\documentclass[preprint%
reprint, 
 amsmath,amssymb,
 aps,superscript address
]{revtex4-2}

\usepackage{graphicx}% Include figure files
\graphicspath{ {./images/}} 
\usepackage{dcolumn}% Align table columns on decimal point
\usepackage{bm}% bold math
\usepackage{verbatim}
\usepackage{hhline}
\usepackage{color}
\usepackage{hyperref} % add hypertext capabilities
\usepackage{ulem}
\usepackage{float}
\hypersetup{
    colorlinks=true,
    linkcolor=blue,
    citecolor=blue,
    urlcolor=black
}
\begin{document}

\preprint{APS/123-QED}

\title{Traveling spatially localized convective structures in an inclined porous medium}% Force line breaks with \\

\author{Zhiwei Dave Li}%
 \email{zwli@uw.edu}
\affiliation{Department of Physics, University of Washington, Seattle, WA 98195, USA}
\author{Chang Liu}%
 \email{chang\_liu@uconn.edu}
\affiliation{School of Mechanical, Aerospace, and Manufacturing Engineering, University of Connecticut, Storrs, CT 06269, USA}
\author{Adrian van Kan}%
 \email{avankan@berkeley.edu}
\affiliation{Department of Physics, University of California at Berkeley, Berkeley, CA 94720, USA}
\author{Edgar Knobloch}%
 \email{knobloch@berkeley.edu}
\affiliation{Department of Physics, University of California at Berkeley, Berkeley, CA 94720, USA}

\date{\today}% It is always \today, today,

\begin{abstract}
The coexistence of multiple stationary, spatially localized structures has recently been reported for convection in an inclined porous layer, but the influence of imperfectly conducting boundaries has not been studied. This paper analyzes the traveling behavior of asymmetric, spatially localized convective structures \textcolor{black}{(`convectons')}, consisting of one or more pulses, in a two-dimensional inclined layer of a porous medium subject to a fixed temperature at the bottom and an imperfectly conducting boundary at the top, such that midplane reflection symmetry is broken. Extensive direct numerical simulations (DNS) are conducted for a wide range of Biot numbers associated with the upper boundary, revealing nontrivial relationships between the drift velocity $c$ of spatially localized structures and the symmetry-breaking parameter $\kappa\geq 0$ based on the Biot number, with $\kappa=0$ corresponding to perfect midplane reflection symmetry associated with $c=0$. In small domains, the drift velocity $c$ is positive (corresponding to upslope motion) and increases monotonically with $\kappa$, while for localized structures in larger domains consisting of a small number of pulses $c$ can be of either sign depending on parameters. For longer structures $c$ reverts to positive values and again increases monotonically with $\kappa$. We show that the along-slope tails of pulses and the associated long-range interactions are governed by the dominant spatial eigenvalues, whose real part is of the smallest magnitude, and we uncover a transition at a finite symmetry-breaking strength $\kappa=\kappa_c>0$: for $\kappa<\kappa_c$, both dominant eigenvalues are complex, implying that the upslope and downslope exponential tails are oscillatory. In contrast, for $\kappa>\kappa_c$, the dominant spatial eigenvalue with a positive real part becomes real, implying that the downslope exponential tail transitions from an oscillatory profile to a monotonic one. As a result, for $\kappa<\kappa_c$, bound states consisting of different numbers of pulses are present. These display rich dynamical phenomena including inelastic collisions leading to other bound states, while for $\kappa\gg\kappa_c$, adjacent pulses are found to repel one another and so tend to spread out, eventually becoming equispaced in the finite computational domain. The strength of the repulsive interaction increases with $\kappa$, i.e., with increased symmetry breaking. A reduced description of this behavior is proposed based on the interaction between tails of adjacent localized structures, which accurately reproduces the repulsion and inelastic collisions observed in DNS. The reduced model indicates that the transition from bound states to an equidistant configuration occurs when the monotonic tail and the oscillatory tail have the same slope, which occurs at a value of $\kappa$ greater than $\kappa_c$. The results presented here represent a first step towards understanding the dynamics of spatially localized patterns in moderate-Rayleigh number convection in an inclined porous medium subject to an imperfectly conducting boundary.
\end{abstract}

\maketitle

\section{Introduction}
Motivated by geophysical and hydrological applications, convection in porous media has been widely studied to analyze carbon dioxide sequestration in saline aquifers \cite{riaz2006onset, wen2018convective}, patterned ground formation \cite{george1989patterned}, the origin of polygonally shaped crusts on salt lakes \cite{lasser2023salt}, and the melting of ice \cite{chang1996natural}. Moreover, the applications of fluid flow in porous media extend beyond geophysics to biomechanics and chemical and mechanical engineering, among others \cite{shenoy1994non}. The study of convection in porous media based on Darcy's law is also known as the B\'enard–Darcy problem, a problem extensively investigated using linear stability analysis \cite{horton1945convection,rogers1953convection}, bifurcation analysis \citep{mamou1998double,mamou1999thermosolutal,mahidjiba2000onset,liu2022single}, and direct numerical simulations in the fully-developed turbulent regime \citep{otero2004high,hewitt2012ultimate,hewitt2014high,wen2015structure,pirozzoli2021towards,zhu2024transport}; see the reviews \citep{nield2013historical,huppert2014fluid,hewitt2020vigorous,de2023convective} and the textbook \citep{nield2006convection} on porous medium convection.

Studies of porous medium convection have evolved from the original setup investigated since the 1940s to incorporate additional effects like anisotropy \citep{storesletten1998effects,ennis2005onset,de2017dissolution}, double diffusion \citep{trevisan1987mass,rosenberg1992thermohalin,mamou1998double,mamou1999thermosolutal,mahidjiba2000onset,neufeld2010convective}, and rotation \citep{vadasz2019instability}, among others. One simple but practically relevant modification, in the context of geological and engineering applications, is to consider porous layers that are inclined with respect to the horizontal. The inclination of the layer leads to a background shear flow, which can significantly modify the convective instability \cite{wen2018inclined,wen2019moderate,reetz2020invariant1,reetz2020invariant2,singh2023longitudinal}.
Experimental and numerical studies have revealed the existence of three distinct flow regimes near the onset of convection in an inclined porous layer: a stable unicellular flow regime and two additional regimes characterized by the presence of polyhedral cells and helicoidal cells, respectively \cite{bories1973natural,caltagirone1985solutions}. 
These studies also examined the transitions between the different flow regimes encountered as the Rayleigh number $Ra$ and the inclination angle $\phi$ are varied. Caltagirone and Bories \cite{caltagirone1985solutions} were the first to domonstrate that the laminar base state is linearly stable for $Ra \cos{\phi}\leq4\pi^2$. In a more recent numerical investigation, Rees and Bassom \cite{rees2000onset} suggested that a small angle $\phi$ can lead to linear instability in a two-dimensional inclined porous layer at large $Ra$, while at a large $\phi$ the base state is linearly stable and the system does not exhibit convection. Subsequent work by Wen and Chini \cite{wen2018inclined} showed that in domains with large aspect ratio the base state is not energy-stable for any angle $0\leq\phi\leq90^{\circ}$ at $Ra > 91.6$, with sufficiently large disturbances yielding distinct forms of convective flows, including stable, stationary, spatially localized convection patterns with various numbers of roll pairs (or `pulses') \cite{wen2019moderate}. 

Spatially localized structures have been found in a wide range of physical systems, both conservative and dissipative \cite{purwins2010dissipative,knobloch2015spatial}. Knobloch \citep{knobloch2015spatial} provides a review of spatially localized structures in different dissipative systems, starting with the Swift-Hohenberg equation \cite{PhysRevE.73.056211,burke2009swift,houghton2011swift} and its relatives, which provide an order-parameter description of the dynamics in many cases, extending the discussion to more complex systems including granular dynamics and fluid flows, among others. For example, spatially localized structures are widely observed in subcritical regimes of wall-bounded shear flows \citep{schneider2010snakes}, and play an important role in the transition to turbulence; see the reviews \citep{graham2021exact,kawahara2012significance}. 

Moreover, spatially localized structures are also widely observed in convectively driven fluid flows and come in different forms. A common type of localized structure found in convective systems is the modulated traveling wave, consisting of a traveling wave state propagating with phase speed $c_p$ under an envelope that is itself moving with a distinct group speed $c_g$. As a result, such states are quasi-periodic; their existence is related to the traveling waves created in a Hopf bifurcation from the conduction state. Such modulated traveling waves have been extensively studied in the context of binary fluid convection, in \textcolor{black}{both experimental \cite{kolodner1989dynamics,steinberg1989pattern,kolodner1991drift} and numerical settings \cite{barten1991localized,barten1995convection,jung2005traveling,watanabe2016skeleton}, and collisions between them have also been investigated \cite{kolodner1991collisions,taraut2012collisions,iima2005collision}}. Rather than modulated traveling waves, we investigate here traveling localized states that are stationary in a comoving frame, also known as \textit{convectons}, a term first coined in the context of magnetoconvection \cite{blanchflower1999magnetohydrodynamic}. In contrast with modulated traveling waves, convectons are born in a steady-state secondary bifurcation from \textcolor{black}{lower branch unstable spatially periodic states arising from a subcritical bifurcation from the base state}. In general, convectons remain stationary if they are left-right symmetric. Traveling motion of convectons therefore requires a broken reflection symmetry in the system, which may be the midplane reflection symmetry, \textcolor{black}{ i.e.,} the Boussinesq symmetry. This is physically different from the case of modulated traveling wave packets, which move because of a non-zero group speed. Convectons (both traveling and stationary) have not only been studied extensively in magnetoconvection \cite{blanchflower1999magnetohydrodynamic,blanchflower2002three,dawes2007localized,jacono2011magnetohydrodynamic}, but also in binary fluid convection, \cite{batiste2005simulations,batiste2006spatially,mercader2008spatiotemporal,lo2010spatially, mercader2011dissipative,mercader2011convectons,mercader2013travelling,jacono2013three}, including in a porous medium \cite{jacono2017complex} and with surface tension effects \cite{assemat2008spatially}, and have also been seen in rotating convection \cite{beaume2013convectons} \textcolor{black}{and in double-diffusive convection \citep{beaume2011homoclinic,beaume2013convectons_double,beaume2018three,tumelty2023toward}.} The transition of a steady state into a traveling one as a result of forced symmetry breaking is widely observed in pattern-forming systems \cite{schutz1995transition,lo2017localized,ophaus2018resting,raja2023collisions}. In the work by Wen and Chini on inclined porous medium convection \citep{wen2019moderate}, fixed temperature boundary conditions were applied at the top and the bottom surfaces, preserving the midplane reflection symmetry and resulting in stationary localized convective structures. In practice, however, it is also physically relevant to consider imperfectly conducting boundary conditions which may lead to symmetry breaking and hence drifting motion. 

In this work, we investigate the traveling behavior of convectons generated by the breaking of midplane reflection symmetry due to differing temperature boundary conditions at the top and bottom boundaries. We start by reproducing the stationary, spatially localized structures, consisting of between one and five pulses, previously found at the moderate Rayleigh number $Ra = 100$, for an inclination angle $\phi = 35^{\circ}$ in domains with aspect ratio $L_x = 10$ and fixed temperature at both top and bottom boundaries \citep{wen2019moderate}. We use these spatially localized structures to initialize direct numerical simulations (DNS) with asymmetric boundary conditions, with fixed temperature at the bottom boundary and an imperfectly conducting top boundary. Traveling spatially localized structures are analyzed using DNS with \textcolor{black}{varying symmetry-breaking strength and varying computational domains with aspect ratios $L_x\in [10,160]$}. Different initial conditions are also considered, including multiple $n$-pulse ($n=1,2,3,4,5$) structures within a single domain, leading to collisions and other nontrivial behavior.  

The remainder of this paper is structured as follows. In the following section, we describe the problem formulation, as well as the numerical method and relevant control parameters. In Sec.~\ref{sec:section3}, we focus on the dynamics of traveling spatially localized structures with one to five pulses, analyzing the relationship between the drift velocity of these structures and a symmetry-breaking parameter. We also demonstrate using DNS that the finite size of the computational domain has a significant impact on the dynamics and specifically on the stability properties of localized states. In Sec.~\ref{sec:section4}, we introduce spatial eigenvalues to predict the growth/decay rate and wavelength of upslope and downslope tails of these spatially localized structures. These theoretical predictions match the results obtained through DNS and successfully predict the transition from oscillatory to monotonic tails, providing clues to the formation of traveling bound states and the interactions between the localized structures in this system. In Sec.~\ref{sec:section5}, we \textcolor{black}{analyze the long-time temporal dynamics of interacting localized structures. We show that} when the midplane reflection symmetry is strongly broken, there is a repulsive interaction between pulses whose strength increases with the degree of symmetry breaking. In contrast, at weak symmetry breaking, stable traveling bound states consisting of $n$ pulses exist, which propagate at different drift velocities for different $n$. This leads to nontrivial collision behavior, including the formation of larger stable bound states and a Newton's cradle-like phenomenon near the onset of repulsion, where a larger structure splits into smaller substructures which subsequently collide in the periodic domain, reproducing the initial structure and leading to a cyclic repetition. In Sec.~\ref{sec:section6} we shed light on aspects of this behavior via a reduced description of the interaction between adjacent localized structures and show that much of the observed behavior is due to the interaction between the trailing tail of the leading structure and the leading tail of the trailing structure. The paper concludes in Sec.~\ref{sec:conclusion} with a discussion of future directions. 

\section{Problem formulation and numerical method}

\subsection{Formulation and control parameters}
\begin{figure}[!htbp]
\includegraphics[width=0.7\linewidth]{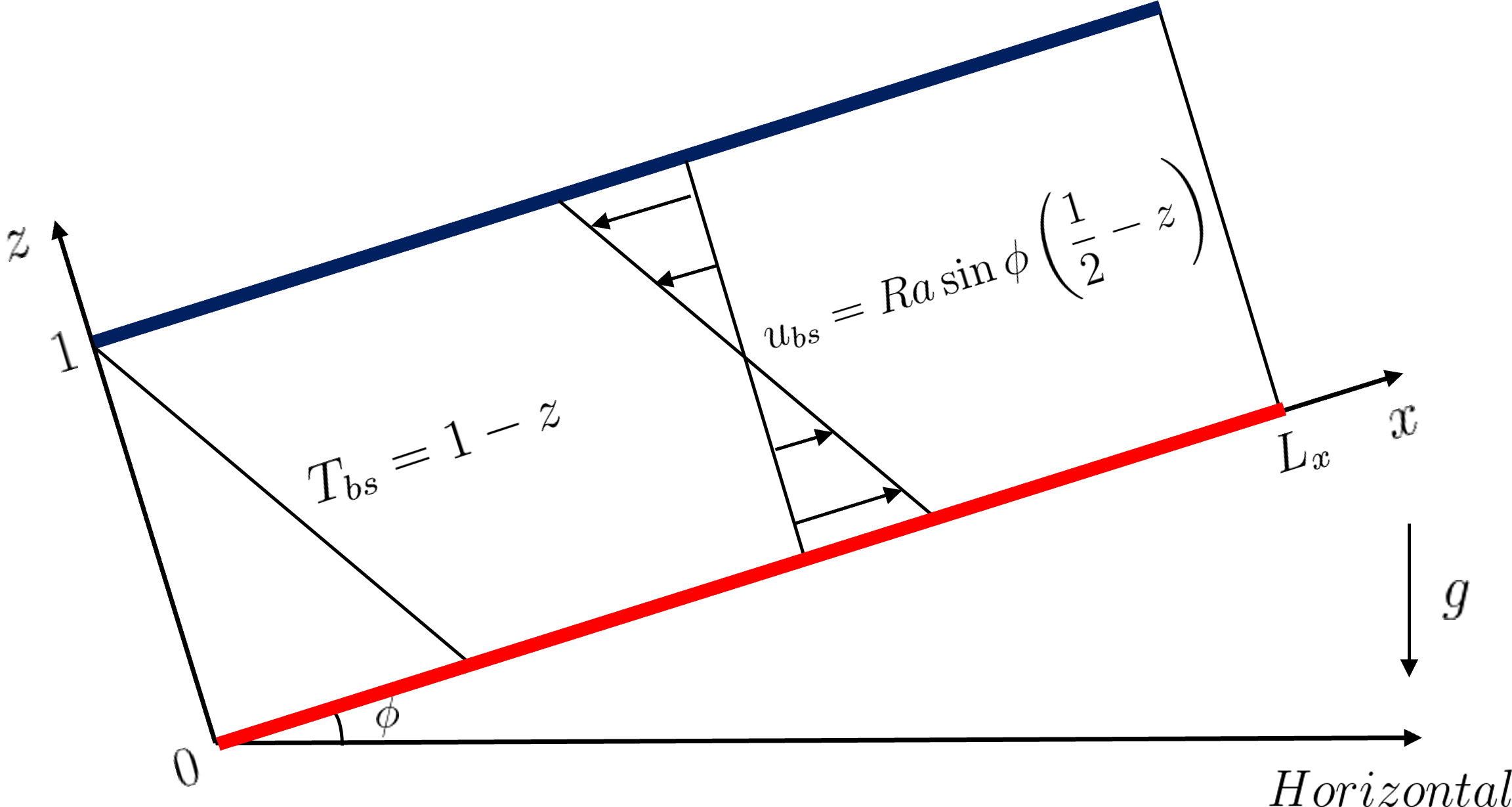}
\caption{Setup and dimensionless base state for 2D convection in a porous Rayleigh-B\'enard cell inclined at an angle $\phi$ to the horizontal. The domain is taken to be periodic in the wall-parallel ($x$) direction. The layer is heated from below ($z=0$) and cooled from above ($z=1$). The base state temperature is $T_{bs}=1-z$ with background shear $\boldsymbol{u}_{bs}=u_{bs} \mathbf{e}_x = Ra\sin\phi\,(\frac{1}{2}-z)\boldsymbol{e}_x$. }
\label{fig:illustration}
\end{figure}

We consider 2D convection in an inclined porous layer with aspect ratio $L_x$ heated from below and cooled from above as shown in Fig.~\ref{fig:illustration}. The system can be described by the following non-dimensional Darcy–Oberbeck–Boussinesq equations \cite{nield2006convection} in terms of velocity $\boldsymbol{u}_{tot} $, temperature $T_{tot}$ and pressure $p_{tot}$:
\begin{subequations}
\begin{align}
\boldsymbol{u}_{tot} + \boldsymbol{\nabla}{p}_{tot} & = Ra\,T_{tot}(\sin\phi\,\boldsymbol{e}_x+\cos\phi\,\boldsymbol{e}_z), \\
\boldsymbol{\nabla}\cdot{\boldsymbol{u}_{tot}} & = 0, \\
\partial_t{T_{tot}}+\boldsymbol{u}_{tot}\cdot\boldsymbol{\nabla}{T_{tot}} & = \nabla^{2}T_{tot},
\end{align}
\end{subequations}
where $\boldsymbol{e}_x$ and $\boldsymbol{e}_z$ are unit vectors in the $x$ (wall-parallel) and $z$ (wall-normal) directions, respectively. We measure lengths in units of the layer height $H$ and time in units of the diffusive time $t_D=H^2/D$, where $D$ is the thermal diffusivity. The velocity is normalized by $H/t_D=D/H$, and the temperature by the background temperature difference $\Delta T$ across the layer. There are three dimensionless control parameters in the problem: the inclination angle $\phi$ of the layer with respect to the horizontal, the domain aspect ratio $L_x\equiv L/H$ and the Rayleigh number
\begin{equation}
Ra \equiv \frac{\alpha g \Delta  T \,KH}{\nu D},
\end{equation}
where $\alpha$ is the thermal expansion coefficient, $g$ is the magnitude of gravitational acceleration, $K$ is the Darcy permeability coefficient, and $\nu$ is the kinematic viscosity.

The inclination of the layer generates a unidirectional shear flow, corresponding to the base state $T_{bs} = 1-z$, $\boldsymbol{u}_{bs} = Ra \sin\phi\left(\frac{1}{2}-z\right)\boldsymbol{e}_x$, and $p_{bs}=\frac{1}{2}Ra\,\sin\phi\,x+Ra\,\cos\phi(z-\frac{1}{2}z^2)$, as shown schematically in Fig.~\ref{fig:illustration}. Decomposing $T_{tot}$, $\boldsymbol{u}_{tot}$, and $p_{tot}$ into the base state and deviations from it, we have 
\begin{subequations}
    \begin{align}
T_{tot} = &T_{bs}+T=  1-z+T, \\
\boldsymbol{u}_{tot} =& \boldsymbol{u}_{bs}+\boldsymbol{u}= Ra \sin\phi\,\left(\frac{1}{2}-z\right)\boldsymbol{e}_x + \boldsymbol{u},\\
p_{tot}=&p_{bs}+p=\frac{1}{2}Ra\,\sin\phi\,x+Ra\,\cos\phi(z-\frac{1}{2}z^2)+p.
\end{align}
\end{subequations}
We can then write the governing equations in terms of the deviation velocity $\boldsymbol{u}$, temperature $T$, and pressure $p$:
\begin{subequations}
\label{eq:perturbation_equations}
    \begin{align}
\boldsymbol{u} + \boldsymbol{\nabla}{p} - Ra\,T(\sin\phi\,\boldsymbol{e}_x+\cos\phi\,\boldsymbol{e}_z) & = 0, \\
\boldsymbol{\nabla}\cdot{\boldsymbol{u}} & = 0, \\
\partial_t{T}+Ra \sin{\phi}\,\left(\frac{1}{2}-z\right)\partial_x{T}-w+\boldsymbol{u}\cdot\boldsymbol{\nabla}{T} & = \nabla^{2}T,
\end{align}
\end{subequations}
where $w$ is the wall-normal component of the velocity perturbation $\boldsymbol{u}$. We apply periodic boundary conditions in the $x$ direction for all variables. For boundary conditions in $z$, we assume impenetrability,
\begin{equation}
    w(z = 0) = w(z = 1) = 0, \label{eq:w_bcs}
\end{equation}
and impose a Dirichlet boundary condition on the temperature {\it deviation} $T$ at the bottom $(z=0)$ and a Robin boundary condition (Newton's law of cooling) at the top $(z=1)$,
\begin{subequations}
\label{eq:T_bcs}
    \begin{align}
    T(z = 0) & = 0, \\
    (1-\kappa)T(z = 1) + \kappa\partial_zT(z = 1) & = 0. 
\end{align}
\end{subequations}
Here, a symmetry-breaking parameter $\kappa \in [0, 1]$ is introduced in the boundary conditions, controlling the breaking of midplane reflection symmetry in terms of the upper temperature boundary condition. We note that the Biot number associated with the upper boundary is given by $(1-\kappa)/\kappa$ and lies in $[0,\infty)$; see, e.g. Chapter 6 of \cite{nield2006convection}. At $\kappa=0$, the system is subject to symmetric Dirichlet boundary conditions [i.e., $T(z=0)=T(z=1)=0$], leading to symmetry of the problem with respect to (diagonal) reflection ${\cal R}:~(x,z)\rightarrow (-x,1-z),~(u,w,T)\rightarrow -(u,w,T)$. 

When $\kappa=0$ and the inclination $\phi$ exceeds $31.3^{\circ}$ the base state is linearly stable for all values of $Ra$ \citep{wen2019moderate}. However, for sufficiently large $Ra$ stable stationary ${\cal R}$-symmetric spatially localized structures can be generated by suitable finite amplitude perturbations, and these therefore coexist with the stable base state. In the present work we follow \citep{wen2019moderate} and fix the inclination at $\phi = 35^{\circ}$ and the Rayleigh number at $Ra = 100$, focusing on the effects of breaking the reflection symmetry ${\cal R}$ via the symmetry-breaking parameter $\kappa$ and the role played by the aspect ratio $L_x$ of the domain. Linear stability analysis shows that the base state is stable for all $\kappa\in [0,1]$ at the selected parameters $\phi=35^\circ$ and $Ra=100$ indicating that all of the spatially localized structures analyzed here coexist with the stable base state. In particular, we show that for $\kappa>0$ all localized structures acquire a nonzero drift speed, and explore the resulting dynamics.

\subsection{Numerical method and simulation setup}
We use the spectral solver Dedalus \cite{burns2020dedalus} to conduct DNS of the fluid equations, solving Eqs.~(\ref{eq:perturbation_equations}) subject to the boundary conditions specified in Eqs.~(\ref{eq:w_bcs}) and (\ref{eq:T_bcs}). The system is discretized spatially using the Fourier spectral method in the $x$ direction and the Chebyshev spectral method in the $z$ direction. \textcolor{black}{A second-order two-stage diagonally implicit Runge-Kutta (DIRK) + explicit Runge-Kutta (ERK) time-stepping scheme (RK222) \cite[section 2.6]{ascher1997implicit} is used, which is a hybrid time-stepping method combining implicit and explicit Runge-Kutta methods.} In addition to the $L_x = 10$ domain considered in \cite{wen2019moderate}, simulations are systematically performed in large periodic domains ranging from $L_x = 10$ to $L_x = 160$ with the corresponding grid point numbers $n_x$ ranging from $256$ up to $4096$, with $n_x$ proportional to $L_x$. In the $z$ direction, we keep the resolution fixed at $n_z=64$ grid points. These horizontally extended domains allow us to understand more easily the generic properties of $n$-pulse $(n=1,2,3,4,5)$ localized structures and the impact of finite domain size on their dynamics. We first choose different initial conditions to reproduce steady spatially localized structures with one to five pulses when $L_x=10$ and $\kappa=0$ \citep{wen2019moderate} and then use these steady spatially localized structures as initial conditions for $\kappa \neq 0$ simulations. For larger domains $\tilde{L}_x>10$, we concatenate the spatially localized structures obtained with $L_x=10$ and $\kappa=0$ with the trivial state $(\boldsymbol{u},T,p)=(\boldsymbol{0},0,0)$
elsewhere, thereby generating an initial condition for simulations on the wider domain $\tilde{L}_x$, still with $\kappa=0$. We let the simulation with such a concatenated initial condition evolve to a steady state, and then use the result as the initial condition for $\kappa\neq 0$ simulations with the corresponding domain size $\tilde{L}_x$.

\section{Characterization of $n$-pulse traveling structures}
\label{sec:section3}

\begin{figure}[!bp]
\includegraphics[width=1\linewidth]{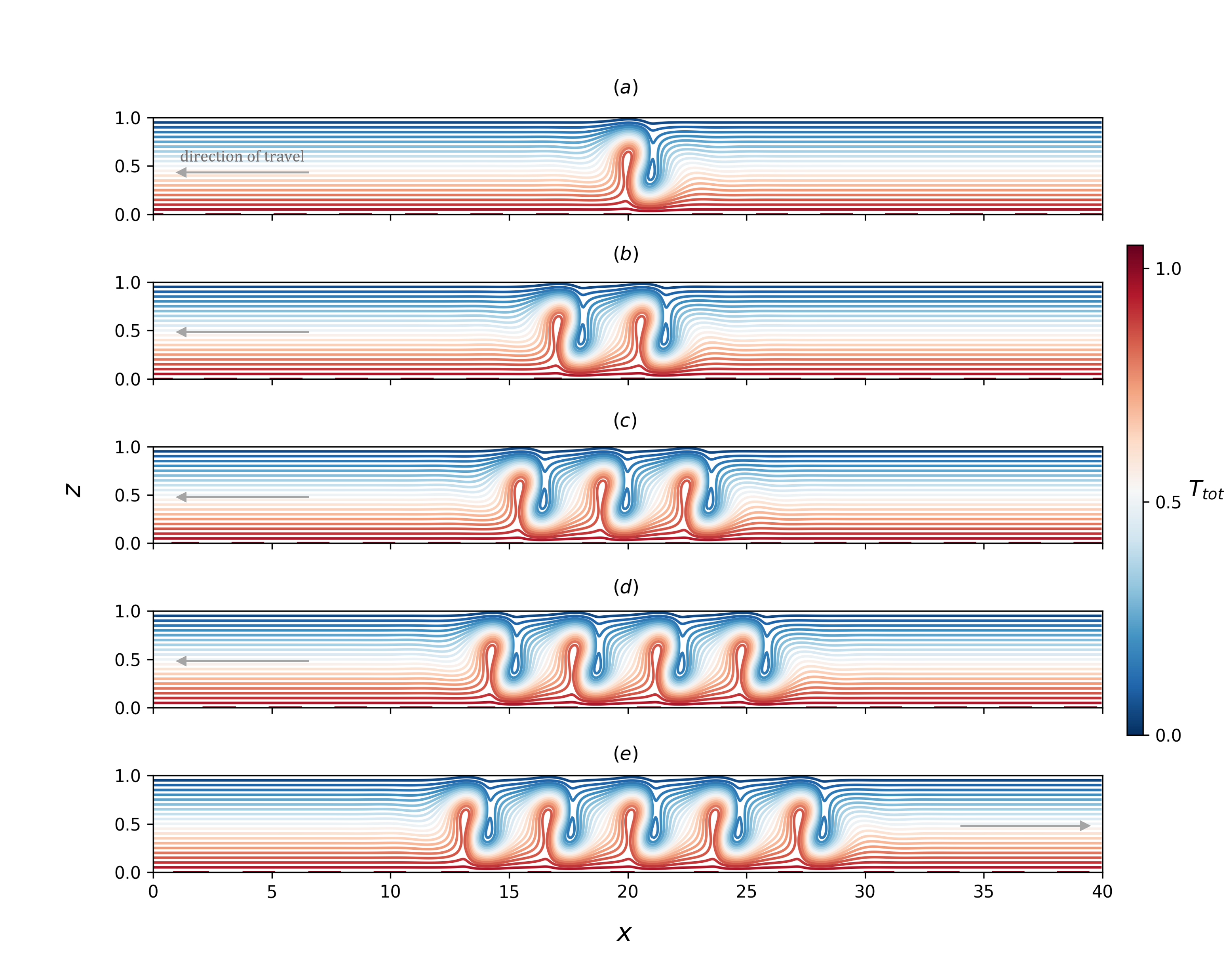}
\caption{Snapshots of the total temperature profile $T_{tot}(x, z)$ of $n$-pulse traveling spatially localized convective structures in a $L_x = 40$ domain, at $\phi = 35^{\circ}$, $Ra=100$, and $\kappa = 0.01$. At this value of $\kappa$, traveling spatially localized structures form bound states. Panels (a)–(e) show spatially localized structures consisting of one to five pulses. }
\label{fig:t2p_0.01}
\end{figure}

\begin{figure}[!tbp]
\includegraphics[width=1\linewidth]{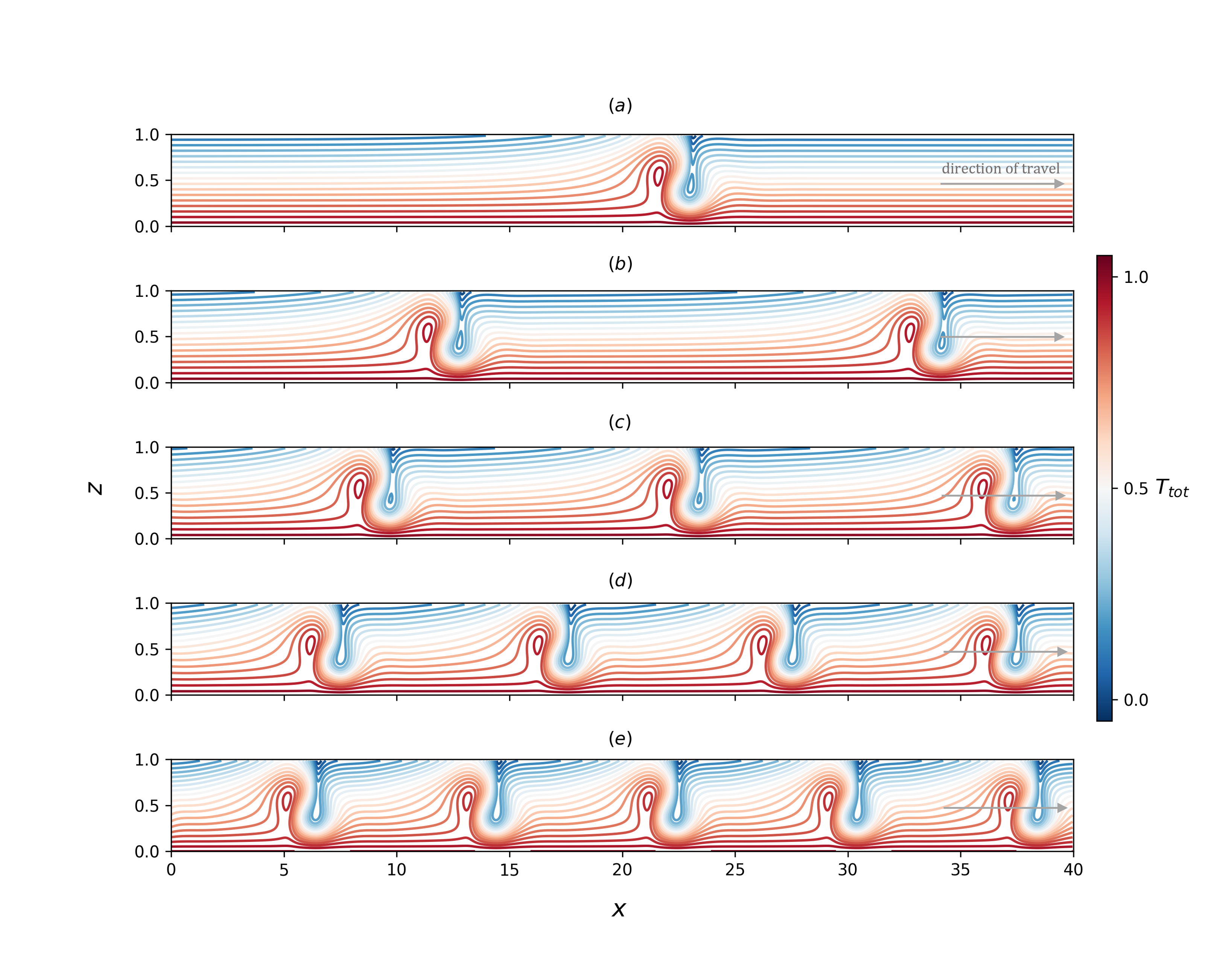}
\caption{Snapshots of the total temperature profile $T_{tot}(x, z)$ of $n$-pulse traveling spatially localized convective structures in a $L_x = 40$ domain, at $\phi = 35^{\circ}$, $Ra=100$, and $\kappa = 0.5$. At this value of $\kappa$, traveling spatially localized structures converge to an equispaced configuration at late times. Panels (a)–(e) show states consisting of one to five pulses.}
\label{fig:t2p_0.5}
\end{figure}

In this section, we focus on the dynamics of traveling localized structures. Extensive DNS were conducted for different combinations of $\kappa$ and $L_x$, generally resulting in $n$-pulse, traveling, spatially localized structures for $\kappa>0$ owing to the breaking of the reflection symmetry ${\cal R}$, cf. \cite{schutz1995transition,lo2017localized,ophaus2018resting,raja2023collisions}. Figures~\ref{fig:t2p_0.01} and \ref{fig:t2p_0.5} show snapshots of the 2D profiles of $n$-pulse traveling convectons with $n=1,2,3,4,5$ at two different values of $\kappa$ in a domain of size $L_x=40$. In each case the structures are shown at late times when they have reached a steady state in the comoving frame, with $\kappa = 0.01$ in Fig.~\ref{fig:t2p_0.01} and $\kappa = 0.5$ in Fig.~\ref{fig:t2p_0.5}. 
When symmetry is only weakly broken ($\kappa=0.01$), pulses in multi-pulse traveling spatially localized structures form bound states with small separation, as shown in Fig.~\ref{fig:t2p_0.01}, hereafter referred to as bound traveling localized structures. Similar bound states of individual pulses, also known as finite pulse trains, are widely observed in many problems featuring spatially localized structures, including reaction-diffusion systems with two \cite{or2000stable} and three species \cite{yochelis2008generation,yochelis2015origin}, and flowing thin films \cite{bergeon2010dynamics}, to name only a few.

At strong symmetry-breaking (i.e. $\kappa\gtrsim 0.1)$, pulses will repel each other, leading to the equispaced states shown in Fig.~\ref{fig:t2p_0.5} at $\kappa=0.5$. The distance between neighboring pulses in this state is equal to the domain size divided by the number of pulses, indicating that they reach the largest possible separation in the given finite domain. In some of the extreme cases where symmetry-breaking is sufficiently strong ($\kappa$ close to unity) or the domain is particularly small, the spatially localized structures are unstable and decay to the base state. In this section we select the domain size and $\kappa$ to exclude this trivial case; the mode of instability of these states will be probed further in Sec.~\ref{sec:section5}. Considering the large number of parameter combinations ($\approx 150$) analyzed in this section, we only consider relatively short DNS with simulations lasting between $t = 500$ and $t = 2000$ diffusive time units, corresponding to approximately $526$ to $1152$ CPU hours for each parameter combination. Long-time DNS results at selected parameters are presented in Sec.~\ref{sec:section5}.

\subsection{Dependence of drift velocity on $\kappa$}

\begin{figure}[h]
\includegraphics[width=1\linewidth]{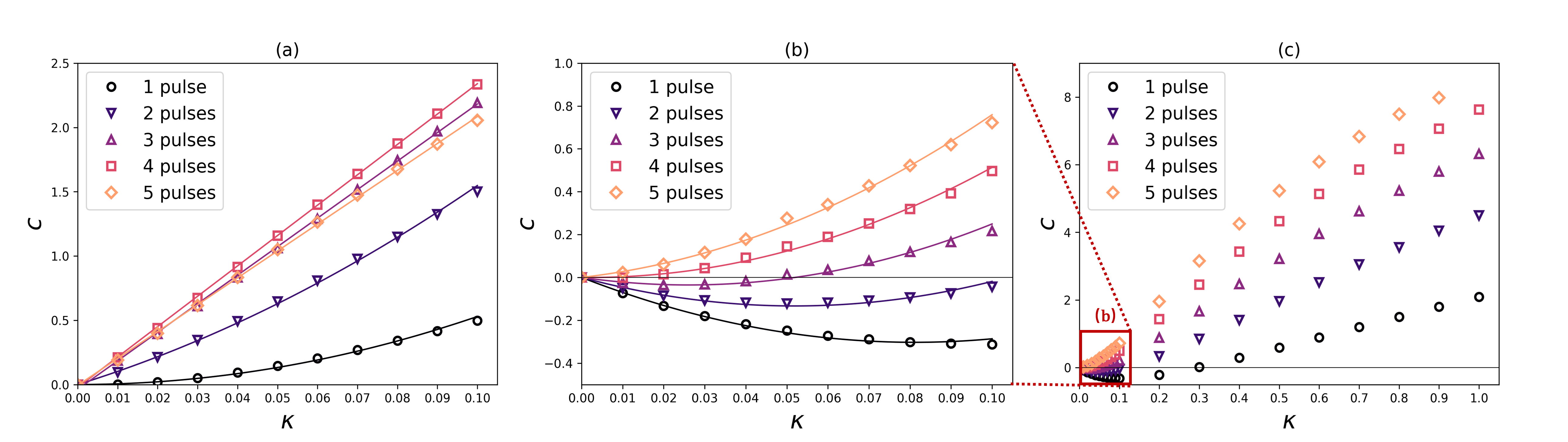}
\caption{Drift velocity $c$ versus the symmetry-breaking parameter $\kappa$ for spatially localized structures consisting of one to five pulses within a domain of size $L_x = 10$ (a) and $L_x = 40$ (b,c). Panel (b) gives a close-up view of panel (c) in the weak symmetry-breaking regime $\kappa\in[0, 0.1]$. For spatially localized structures with more than one pulse, the velocities of individual pulses are measured from DNS, and the average of these velocities is taken as the drift velocity $c$ and plotted in markers. In panel (a), $c(\kappa)$ for single-pulse and two-pulse spatially localized structures are fitted with quadratic curves, while those for three- to five-pulse structures are fitted with linear curves. In panel (b), $c(\kappa)$ for single- to five-pulse structures is fitted with quadratic curves. Note that in panel (c) the velocity of the five-pulse spatially localized structure at $\kappa = 1$ is absent because the structure is unstable and spontaneously decays into a four-pulse spatially localized structure. Panels (b) and (c) further indicate that in the domain of size $L_x=40$, the velocity $c$ changes sign as $\kappa$ is increased for one, two, three or four pulses ($c<0$ at $\kappa=0.01$ for four pulses), with downslope motion at small $\kappa$ and upslope motion for larger $\kappa$.}
\label{fig:v-kplots}
\end{figure}

Here, we investigate the dependence of the drift velocity $c$ of spatially localized states on $\kappa$ in the range of $\kappa\in [0, 1]$ and focus specifically on the range $\kappa \in [0, 0.1]$, where the symmetry is only weakly broken, for domain aspect ratios $L_x = 10$ and $L_x = 40$. Figure~\ref{fig:v-kplots}(a) shows the drift velocities $c$ measured in different simulations of traveling spatially localized structures versus the symmetry-breaking parameter $\kappa$ in a $L_x = 10$ domain. Figures~\ref{fig:v-kplots}(b) and \ref{fig:v-kplots}(c) are both for $L_x = 40$, with Fig.~\ref{fig:v-kplots}(b) showing a zoom on the interval $\kappa\in [0,0.1]$. We measure the velocity of the traveling spatially localized structures by tracking the $x$ position associated with the maximum temperature deviation $T$ at the midplane ($z = 0.5$) as a function of time. We then fit the discrete spatiotemporal data for constant velocity pulses with a linear function and use the fitted slope to define the drift velocity of each pulse. For structures with multiple pulses, we use the average velocity of all pulses composing the structure to represent the drift velocity of the entire spatially localized structure.

\textcolor{black}{Based on the DNS results in Fig.~\ref{fig:v-kplots}(a) we conclude that when $L_x=10$ the} drift velocities $c$ of $n$-pulse spatially localized structures are always positive and increase monotonically with the symmetry-breaking parameter $\kappa$. Here, the one- and two-pulse structures display a nonlinear law for the curve $c(\kappa)$, while localized structures with three to five pulses display a linear relation $c\propto \kappa$ (solid lines show fits). At a given value of $\kappa$, the relation between drift velocity and the number of pulses is also nontrivial. For example, Fig.~\ref{fig:v-kplots}(a) shows that for between one and four pulses, a larger number of pulses leads to a larger traveling speed $c$, while the five-pulse structure (where the pulses are deformed (not shown) due to finite-size effects in this small domain) travels more slowly than the three- and four-pulse structures. 

Next, we investigate whether these relations between $c$ and $\kappa$ are robust in a larger domain. Figure~\ref{fig:v-kplots}(b) shows the drift velocity $c$ over the same $\kappa$ range $\kappa\in[0, 0.1]$ in the larger domain $L_x=40$. The figure displays a quadratic relation between $c$ and $\kappa$ for all localized structures regardless of the number of pulses. Note that the drift velocity $c$ is sign-indefinite for one, two, three and four pulses, meaning that these localized structures travel in the $-\boldsymbol{e}_x$ (downslope) direction when $\kappa$ is small, but reverse their travel direction as $\kappa$ increases. In contrast to the non-monotonic relation between the drift velocity and the number of pulses in Fig.~\ref{fig:v-kplots}(a) for $L_x=10$, spatially localized structures with more pulses travel at a larger (signed) velocity $c$ when $L_x=40$, given the same boundary conditions. Figure~\ref{fig:v-kplots}(c) shows the full  $\kappa$ range with $\kappa\in[0,1]$, including the regime of stronger symmetry breaking, for $L_x=40$. The drift velocities are seen to increase monotonically with $\kappa$ for $\kappa\gtrsim 0.1$, and the $c(\kappa)$ curves deviate from the simple quadratic form valid for small $\kappa$. In conclusion, the relation between the drift velocity of traveling spatially localized structures and the symmetry-breaking parameter $\kappa$ is nontrivial and generically nonlinear. In addition, a comparison between the results from DNS in the computational domain $L_x = 10$ and $L_x = 40$ suggests that the domain size $L_x$ has a strong impact on the dynamics of the traveling spatially localized structures studied here. 

\subsection{Dependence of drift velocity on domain size $L_x$}

\begin{figure}[!h]
\includegraphics[width=1\linewidth]{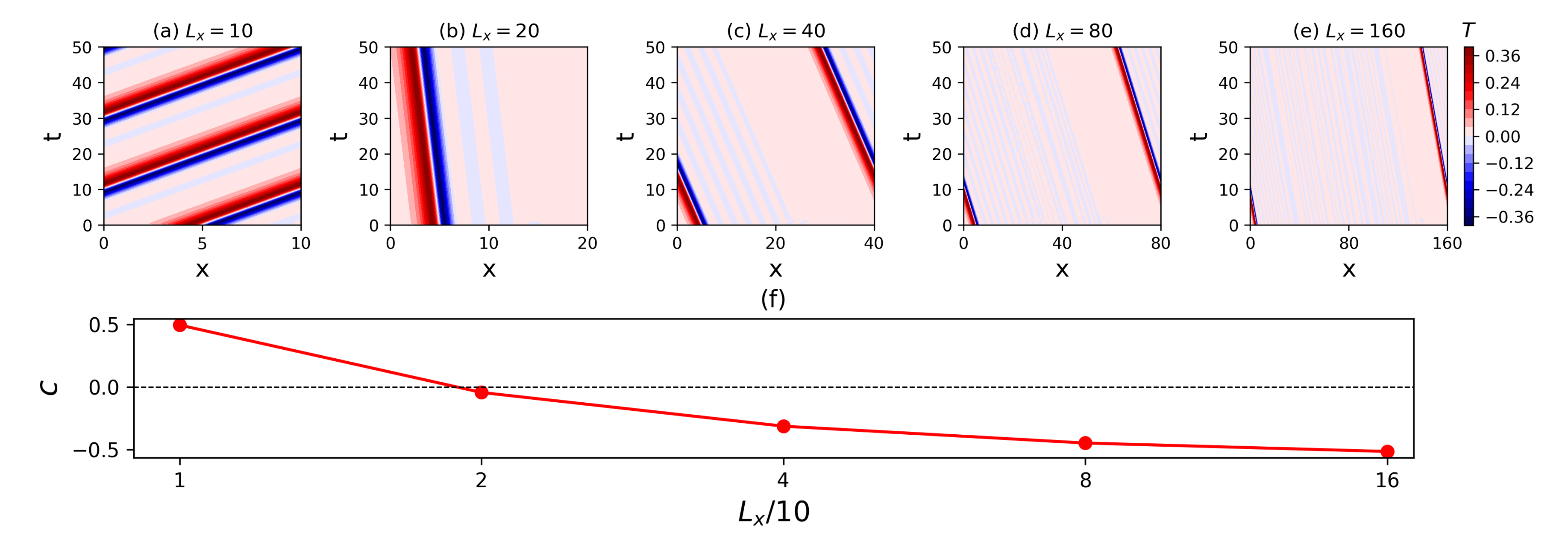}
\caption{Panels (a)-(e) show space-time plots of the temperature deviation $T(x,z=0.5,t)$ at the midplane $z=0.5$ of a single-pulse spatially localized structure traveling in domains of different aspect ratios $L_x$ at $\kappa = 0.1$. Panel (f) shows the drift velocity $c$ as a function of the domain aspect ratio $L_x$. The horizontal axis of panel (f) represents $0.1 L_x$ and is plotted on a logarithmic scale of base 2, so that the points represent $L_x = 10, L_x = 20, L_x = 40, L_x = 80$, and $L_x = 160$.}
\label{fig:5pvlx3}
\end{figure}

Next, we investigate the relation between drift velocity and domain size in more detail. We increased the aspect ratio from $L_x = 10$ to $L_x = 160$ by factors of $2$  and varied the  resolution from $n_x = 512$ to $n_x = 4096$ proportionally. Figure~\ref{fig:5pvlx3} shows a collection of space-time plots of a single-pulse structure traveling in domains of different sizes, from $L_x = 10$ to $L_x = 160$. Colors represent the temperature deviation at the midplane $z=0.5$ so that the large-amplitude, inclined stripes in the space-time plots indicate travel. In the smallest domain, $L_x = 10$, the spatially localized structure travels in the $\boldsymbol{e}_x$ (upslope) direction and therefore has a positive drift velocity. When the domain size is increased to $L_x = 20$, the direction of travel reverses and the localized structure travels in the $-\boldsymbol{e}_x$ (downslope) direction associated with a small, but negative drift velocity. As the domain size increases, the signed drift velocity decreases monotonically as shown in Fig.~\ref{fig:5pvlx3}(f), appearing to saturate at an asymptotic value for large domain sizes. 

Different dynamics originating from domain size differences can be attributed to the interactions between traveling spatially localized structures and their periodic copies. The repulsive interaction between periodic copies resembles the interaction of adjacent pulses in multi-pulse traveling spatially localized structures in DNS. For example, the single pulse travels at $c\approx0.4968$ for $\kappa=0.1$ and $L_x=10$ in Fig.~\ref{fig:5pvlx3}(a), which is the same value of $c$ (to three decimal places) as for the equispaced four-pulse structures with $\kappa=0.1$ and $L_x=40$ in Fig.~\ref{fig:v-kplots}(b). Moreover, the trend in Fig.~\ref{fig:5pvlx3}(f) that the signed drift velocity decreases with increasing domain size $L_x$ is also consistent with the trend in Fig.~\ref{fig:v-kplots}(b) and Fig.~\ref{fig:v-kplots}(c) that the signed drift velocity decreases as the number of pulses is decreased for a given domain size, leading to wider spacing between pulses. 

\subsection{Unstable states at large $\kappa$}

Our DNS results suggest that traveling spatially localized structures can be unstable for certain domain sizes $L_x$ and symmetry-breaking parameters $\kappa$. This instability of traveling localized structures is typically associated with strong symmetry breaking (i.e., large $\kappa$) or small spacing of pulses due to a large number of pulses for a given domain size. A combination of both factors greatly increases the likelihood of instability.  
For example, DNS in a $L_x = 10$ domain with $\kappa \geq 0.2$ reveal that a five-pulse state is unstable,  leading us to focus on the range $\kappa\in [0,0.1]$ in Fig.~\ref{fig:v-kplots}(a), where such structures are stable. Increasing the domain size can suppress this instability. For example, the unstable five-pulse structure at $\kappa = 0.2$, $L_x=10$ is stable at the same value of $\kappa$ in a $L_x=40$ domain; the drift speed of this state is plotted in Fig.~\ref{fig:v-kplots}(c). Nevertheless, a sufficiently strong symmetry breaking in the boundary conditions can generate instability even in a larger domain. For example, at $\kappa = 1$, instability does occur in the case of a five-pulse structure within a domain of aspect ratio $L_x = 40$. For this reason, no velocity measurement is shown in Fig.~\ref{fig:v-kplots}(c) for the five-pulse structure at $\kappa = 1$, as no such stable structure exists at this value of $\kappa$. Among the cases displaying instability, there are two scenarios for the subsequent evolution: 
\begin{itemize}
  \item Case 1. Partial annihilation: The $n$-pulse traveling spatially localized structure spontaneously decays, after a finite time, into a traveling localized structure consisting of fewer pulses. In this process, \textcolor{black}{one or more pulses in the multi-pulse structure decay. In an $L_x = 10$ domain such partial annihilation was observed at (a) $\kappa = 0.45$ where a four-pulse structure decays into a three-pulse structure, (b) $\kappa = 0.5$ and $\kappa=0.6$ where a four-pulse structure decays into a two-pulse structure, and (c) $\kappa = 0.7$ and $\kappa=0.8$ where a four-pulse structure decays into a single traveling pulse. Moreover, in a $L_x = 40$ domain with (d) $\kappa = 1$, a five-pulse structure decays into a four-pulse structure.}
 \item Case 2. Spontaneous (complete) decay: The localized state cannot maintain its structure in the presence of sufficiently strong asymmetry in the boundary conditions and collapses \textcolor{black}{to the base state. In an $L_x=10$ domain such spontaneous decay was observed in the parameter interval (a) $\kappa \in [0.2,1]$ within which five-pulse structures collapse to the base state, and (b) $\kappa\in [0.85,1]$ within which four-pulse structures collapse to the base state.}
\end{itemize}

\begin{figure}[!t]
\includegraphics[width=0.85\linewidth]{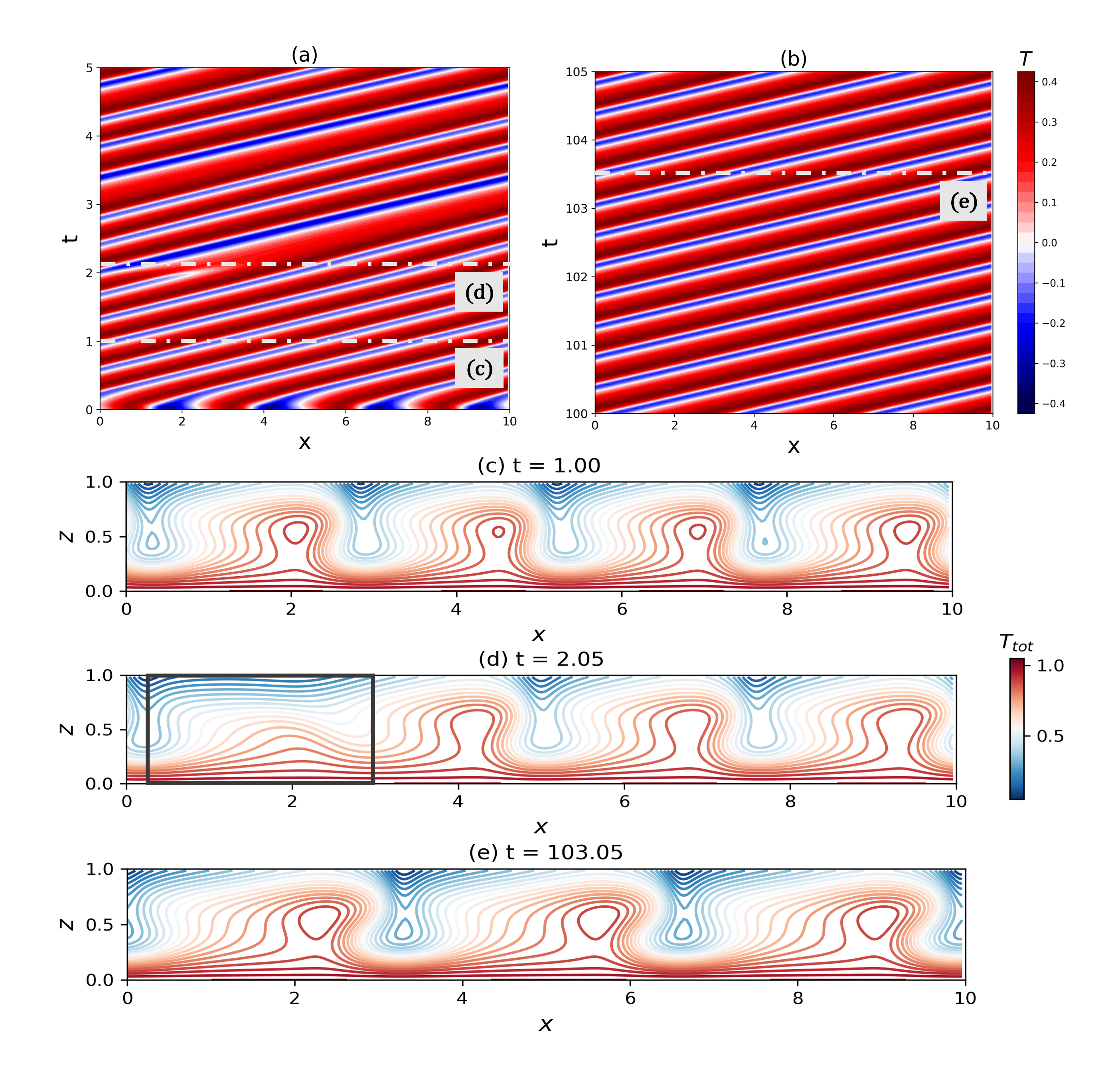}
\caption{Panels (a) and (b) show space-time plots of the temperature deviation $T(x, z=0.5,t)$ at the midplane $z=0.5$ of a traveling four-pulse localized structure in a $L_x = 10$ domain at $\kappa = 0.45$. Panels (c)-(e) show the total temperature profile $T_{tot}(x, z)$ at $t=1$, $t=2.05$, and $t=103.05$, respectively, illustrating the initial stages of the transition from a four-pulse state to a three-pulse state via partial pulse annihilation indicated by the horizontal lines in panels (a) and (b). The black box in panel (d) highlights the collapsing pulse at $t = 2.05$.} 
\label{fig:annilation}
\end{figure}

Panels (a) and (b) of Figures~\ref{fig:annilation}(1) and \ref{fig:annilation}(b) show space-time plots and one-dimensional(1D) temperature profiles representing spatially localized structures at different stages of a transition from a four-pulse state to a three-pulse state described in Case 1 \textcolor{black}{with $L_x=10$ and $\kappa=0.45$}. A symmetric four-pulse initial state at $t=0$ in a $L_x = 10$ domain rapidly becomes asymmetric owing to the finite value of $\kappa$ [$\kappa = 0.45$, Fig.~\ref{fig:annilation}(c)] after which one of the constituent pulses suddenly collapses as can be seen in the space-time plot in Fig.~\ref{fig:annilation}(a) at $t\approx 2.05$. This is also shown in the snapshot of the 2D temperature profile in Fig.~\ref{fig:annilation}(d), where the leftmost pulse is seen to be decaying. As a result, one of the adjacent pulses expands into the void left by the decaying pulse.
However, the resulting asymmetric state, with two narrow pulses and one wider pulse, is transient and gradually reorganizes itself into the regular three-pulse structure with three identical pulses over time, while it continues to travel, with the pulses deforming and spreading out in the domain. Figure~\ref{fig:annilation}(e) explicitly shows that at a later time, all three pulses have become identical in shape and reached an equispaced configuration, identical to the final state observed at $\kappa = 0.45$ when initialized with a three-pulse state (from $\kappa=0$). Similar partial annihilation of the traveling structure is also observed in a simulation initialized with a four-pulse state in an $L_x = 10$ domain at $\kappa = 0.5$, where two out of four pulses decay while traveling and the four-pulse structure evolves into a two-pulse final state. It is interesting to note that the drift velocity $c$ of the ephemeral four-pulse structure prior to the collapse of the fourth pulse ($t\approx 1$) is very close to that of the three-pulse state at late time ($t\approx 100$). The spontaneous decay in Case 2 \textcolor{black}{(not shown)} likely indicates the presence of a fold where an $n$-pulse state ceases to exist, resulting in the collapse of the solution to the trivial state. Such behavior is characteristic of a subcritical state, an observation consistent with the results of \cite{wen2019moderate}.

\section{Spatial eigenvalues and spatially localized structures}
\label{sec:section4}

In the previous sections, we encountered the phenomenon of repulsion between neighboring pulses at strong symmetry breaking, which was shown in some cases to lead to equispaced, domain-filling structures. To elucidate the underlying mechanism of such a repulsive interaction, we analyze here the tail profiles of traveling spatially localized structures using spatial eigenvalues. 

\subsection{Spatial eigenvalues}

Here, we linearize the equations of motion about the base state $(T_{bs}, \boldsymbol{u}_{bs},p_{bs})$, and employ the notion of spatial dynamics \cite{kirchgassner1982wave, haragus2011local} to characterize the tail profiles of spatially localized structures in terms of spatial eigenvalues which have been extensively used to characterize spatially localized structures \cite{kirchgassner1982wave,burke2008classification, haragus2011local,burke2012localized,knobloch2015spatial,parra2018bifurcation,knobloch2021stationary,parra2021origin,verschueren2021dissecting,frohoff2021localized,al2021localized}, including in the Lugiato–Lefever equation \cite{parra2018bifurcation} and in the Swift-Hohenberg equation \cite{PhysRevE.73.056211,raja2023collisions}. 

For the 2D flow considered here, we can express velocity in terms of a stream function $\psi$, so that $u = \partial_z\psi$ and $w = -\partial_x\psi$. Assuming a traveling-wave type solution of the form $\psi(x-ct)$, $T(x-ct)$ with a constant drift velocity $c$, we can write the following scalar governing equations, linearized around the base state $(T_{bs}, \boldsymbol{u}_{bs})$, as
\begin{subequations}
    \begin{align}
    (\partial_x^2+\partial_z^2)\psi & = Ra\,(\partial_z T\sin\phi-\partial_xT\cos\phi),\label{eq:linear_a} \\
    %\partial_tT
    -c\, \partial_xT +Ra\sin\phi\, \left(\frac{1}{2}-z\right)\partial_xT+\partial_x\psi & =(\partial_x^2+\partial_z^2)T. \label{eq:linear_b}
\end{align}
\end{subequations}
Reorganizing equations \eqref{eq:linear_a} and \eqref{eq:linear_b} into a spatial dynamics form yields
\begin{equation}
\partial_x \begin{bmatrix} 
\psi \\ \partial_x\psi \\ T \\ \partial_x T
\end{bmatrix}
= \begin{bmatrix}
0 & 1 & 0 & 0 \\
-\partial_z^2 & 0 & Ra\sin{\phi}\,\partial_z & -Ra \cos{\phi} \\
0 & 0 & 0 & 1 \\
0 & 1 & -\partial_z^2 & -c+Ra\sin{\phi}\,\left(\frac{1}{2}-z\right)
\end{bmatrix}
\begin{bmatrix}
\psi \\ \partial_x\psi \\ T \\ \partial_x T
\end{bmatrix}
\label{eq:spatial_eigenvalue}
\end{equation}
with $x$ playing the role of a time-like variable.
The boundary conditions at the top $(z=1)$ and the bottom ($z=0$) surfaces are
\begin{subequations}
\label{eq:bcs}
    \begin{align}
\partial_x\psi(z=0) = \partial_x\psi(z=1) = 0, \\
T(z=0) = 0, \\
(1-\kappa)T(z=1) + \kappa\partial_zT(z=1) = 0.
\end{align} 
\end{subequations}
We assume that the solution of this linear boundary value system takes the form 
\begin{align}
\begin{bmatrix}
\psi \\ T
\end{bmatrix} = \begin{bmatrix}
\hat{\psi}(z) \\ \hat{T}(z) 
\end{bmatrix} e^{\lambda x} + c.c.,
\end{align}
where $\lambda$ is a spatial eigenvalue that can be obtained by solving Eq.~\eqref{eq:spatial_eigenvalue} together with \eqref{eq:bcs}, and c.c. denotes the complex conjugate. We discretize the $\partial_z$ operators in \eqref{eq:spatial_eigenvalue} using the Chebyshev differentiation matrix \citep{weideman2000matlab}, and then solve for the spatial eigenvalues. Note that the drift velocity $c$ of the traveling localized structure is needed in Eq.~\eqref{eq:spatial_eigenvalue}, but solving for $c$ analytically remains challenging. For this reason, we resort to DNS to measure the drift velocity $c$ of a single pulse traveling in a domain of aspect ratio $L_x = 160$ with the corresponding $\kappa$ as done in Sec.~\ref{sec:section3} and substitute this value into the spatial eigenvalue computation in Eq.~\eqref{eq:spatial_eigenvalue}.

In general, one expects the spatial profile of a pulse to approach the trivial state along the eigenvector corresponding to the spatial eigenvalue with the smallest magnitude of the real part, \textcolor{black}{the contributions from the other eigenvalues having decayed earlier.} In the following, we refer to such eigenvalues as dominant. The dominant spatial eigenvalues thus determine the profile of small-amplitude, exponential tails of the localized structures \cite{knobloch2021stationary}. Real spatial eigenvalues indicate a monotonic tail, while complex eigenvalues indicate an oscillatory tail with the real part of the dominant spatial eigenvalue determining the spatial growth/decay rate of the exponential tail, while its imaginary part determines the wave number of the oscillatory tail. 
In the following, a tail is called `upslope' if its amplitude decreases with $\boldsymbol{e}_x$, behavior associated with dominant spatial eigenvalues whose real part is negative. We denote a tail as `downslope' if its amplitude decreases in the $-\boldsymbol{e}_x$ direction (i.e., increases in the $\boldsymbol{e}_x$ direction), corresponding to dominant spatial eigenvalues with positive real part. 

\begin{figure}[!tbp]
\includegraphics[width=1\linewidth]{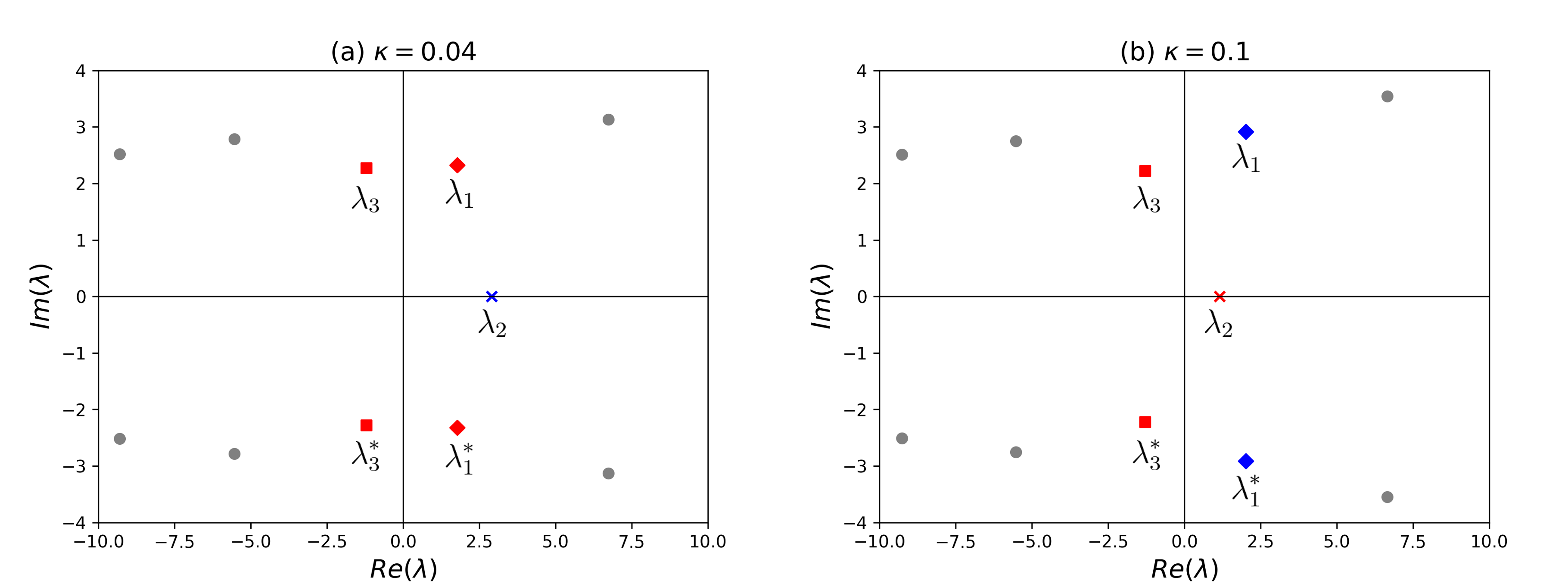}
\caption{The spatial eigenvalue spectrum shown in terms of the real and imaginary parts of the first 11 leading spatial eigenvalues near the origin for the traveling single-pulse localized structure in a domain of size $L_x=160$ at (a) $\kappa=0.04$ and (b) $\kappa = 0.1$. Red markers indicate the dominant spatial eigenvalues in each case. The alternation in color between red and blue for $\lambda_1$ (diamond markers) and $\lambda_2$ (cross markers) suggests a competition between real and complex spatial eigenvalues, which determines whether the downslope tail is monotonic or oscillatory. Meanwhile, $\lambda_3$ (square markers) remains dominant and determines the profile of the upslope tail. } 
\label{fig:spaeigcomp}
\end{figure}

Figure~\ref{fig:spaeigcomp} shows the spectrum of the spatial eigenvalues at $Ra = 100, \phi = 35^{\circ}$ with the symmetry-breaking parameter (a) $\kappa = 0.04$ and (b) $\kappa=0.1$. In Fig.~\ref{fig:spaeigcomp}(a), the dominant spatial eigenvalues consist of two pairs of complex conjugate eigenvalues $\lambda_1, \lambda_1^*$ (with positive real part) and $\lambda_3, \lambda_3^*$ (with negative real part). In Fig.~\ref{fig:spaeigcomp}(b) with stronger symmetry breaking, the dominant spatial eigenvalues consist of one real eigenvalue $\lambda_2>0$ and a pair of complex conjugate eigenvalues $\lambda_3$ and $\lambda_3^*$ with negative real part. Thus the upslope tail is always oscillatory while the downslope tail may be monotonic or oscillatory, depending on whether $\lambda_2<Re(\lambda_1)$ or vice versa.

\begin{figure}[!htbp]
\includegraphics[width=1\linewidth]{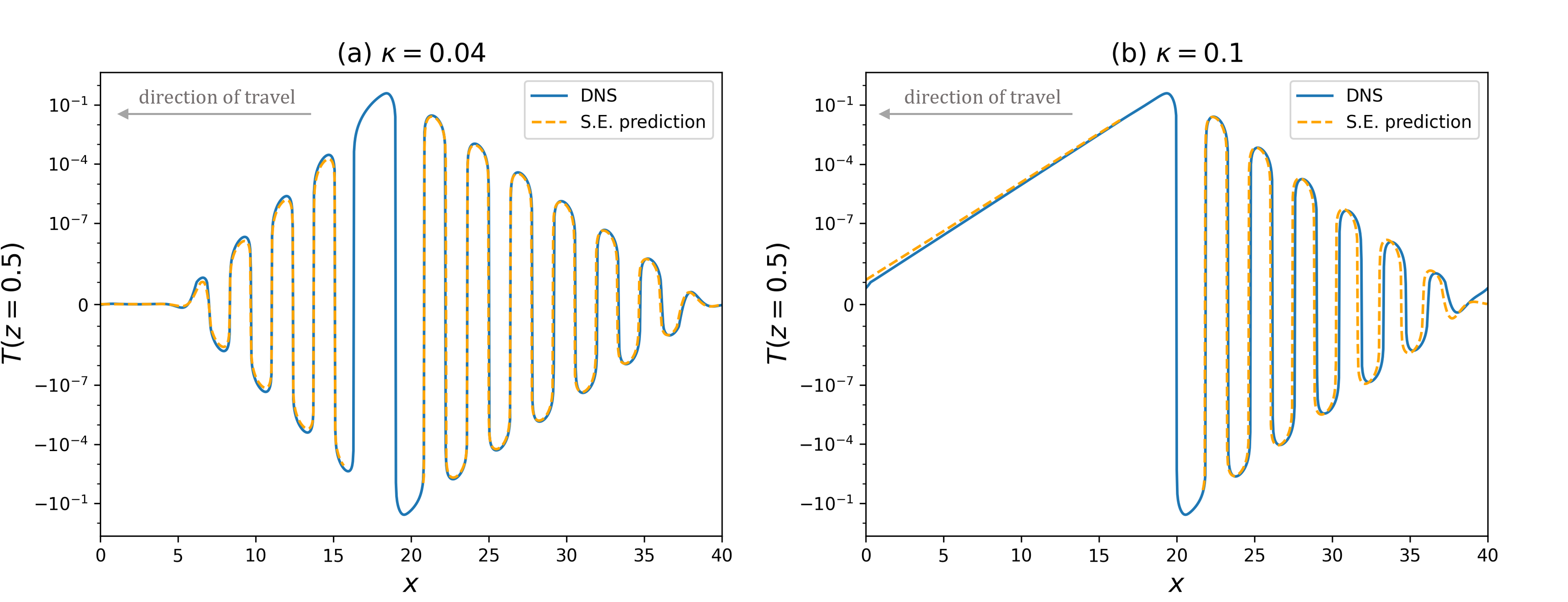}
\caption{Midplane temperature profile $T(x, z=0.5)$ of traveling single-pulse spatially localized structure in a $L_x = 40$ domain as obtained from DNS at (a) $\kappa = 0.04$ and (b) $\kappa = 0.1$. The vertical axis is plotted on a symmetric logarithmic scale, with a cut-off at $ T(x,z=0.5)=\pm 10^{-10}$. The orange dashed lines are the leading and trailing tails as predicted by the dominant spatial eigenvalues in Fig.~\ref{fig:spaeigcomp}.}
\label{fig:spaeigfit}
\end{figure}

To validate the predictions of the tail structure based on the spatial eigenvalue computations discussed above, we performed DNS of single-pulse localized structures traveling in a $L_x = 40$ domain, at (a) $\kappa = 0.04$ and (b) $\kappa = 0.1$. Their 1D temperature deviation profiles in the midplane $T(x,z=0.5)$ are shown in panels (a) and (b) of Fig.~\ref{fig:spaeigfit}, respectively. At $\kappa = 0.04$, both the downslope and upslope tails are oscillatory, corresponding to complex dominant spatial eigenvalues $\lambda_1, \lambda_1^*$ and $\lambda_3, \lambda_3^*$ in Fig.~\ref{fig:spaeigcomp}(a). As expected, at $\kappa = 0.1$, the downslope profile is monotonic, corresponding to the real dominant spatial eigenvalue $\lambda_2>0$, while the upslope profile remains oscillatory, in agreement with the complex spatial eigenvalues $\lambda_3, \lambda_3^*$ shown in Fig.~\ref{fig:spaeigcomp}(b). As the localized structures are traveling downslope ($-\boldsymbol{e}_x$ direction) in Fig.~\ref{fig:spaeigfit}, here the downslope tail can also be referred to as the leading tail while the upslope tail is the trailing tail. 

Figure~\ref{fig:spaeigfit} compares the temperature profiles obtained in DNS with the tail structure predicted from the spatial eigenvalues. In the case of oscillatory tails, we present the spatial eigenvalue prediction as a function of the form $T(x, z=0.5) = \gamma e^{{\rm Re}(\lambda_i) x} \cos({\rm Im}(\lambda_i) x+\theta)$, where $\lambda_i$ ($i=1,3$) is the dominant complex spatial eigenvalue, and the parameters $\gamma$ and $\theta$ are determined by fitting this function to the DNS results. In the case of monotonic tails, a function of the form $T(x, z=0.5) = \alpha e^{\lambda_2 x}$ is used instead, where $\lambda_2$ is the dominant real spatial eigenvalue and $\alpha$ is determined by fitting. As shown in Fig.~\ref{fig:spaeigfit}, the midplane temperature profiles predicted by the spatial eigenvalues agree well with profiles obtained from the DNS results. We mention that hybrid configurations with both monotonic and oscillatory exponential tails, similar to those found here for $\kappa>\kappa_c$, have also been found in other systems, including the Kuramoto-Sivashinsky equation \cite{kawahara1988pulse} describing a falling liquid film \cite{kalliadasis2007thin}.

Furthermore, to quantitatively validate the spatial eigenvalue prediction, we can measure the spatial growth/decay rate of both oscillatory and monotonic tails and measure the wave number of the oscillatory tail by fitting the DNS result.
For the oscillatory tail, we measure the average separation between local maxima representing wavelength $\ell$ leading to wave number $k = \frac{2\pi}{\ell}$. The growth/decay rates $\beta$ of the oscillatory tails are obtained by fitting the discrete set of these local maxima to $T(x,z=0.5)=\alpha e^{\beta_i x}$ ($i=1,3$). The downslope tails are characterized by $\beta_1>0$, while $\beta_3<0$ in the upslope tail, cf. Fig.~\ref{fig:spaeigfit}. The monotonic tail is directly fitted with the function $T(x,z=0.5)=\alpha e^{\beta_2 x}$ to obtain the spatial growth rate $\beta_2>0$ of the downslope tail.

Tables \ref{tab:table1} and \ref{tab:table2} compare quantitatively the spatial eigenvalue predictions against the spatial growth rate and wave number obtained from DNS of a single-pulse structure at $\kappa=0.2$ in domains of various sizes by the fitting procedure described above. Here, we can see that the spatial eigenvalue prediction approaches the DNS results, with very good agreement for larger domain sizes, where the fitted DNS results become independent of the domain size. \textcolor{black}{In particular, $\beta_2$ and $\beta_3$ in Tables \ref{tab:table1} and \ref{tab:table2} are almost independent of $L_x$ when the domain size $L_x\geq 40$, for which the tail profiles have decayed to machine precision (double precision here). For example, we find $e^{\beta_3L_x}=1.25\times 10^{-47}$ and $e^{-\beta_2L_x}=1.29\times 10^{-22}$ when $\beta_3=-1.35$, $\beta_2=0.63$, and $L_x=80$.}

\begin{figure}[!htbp]
\includegraphics[width=1\linewidth]{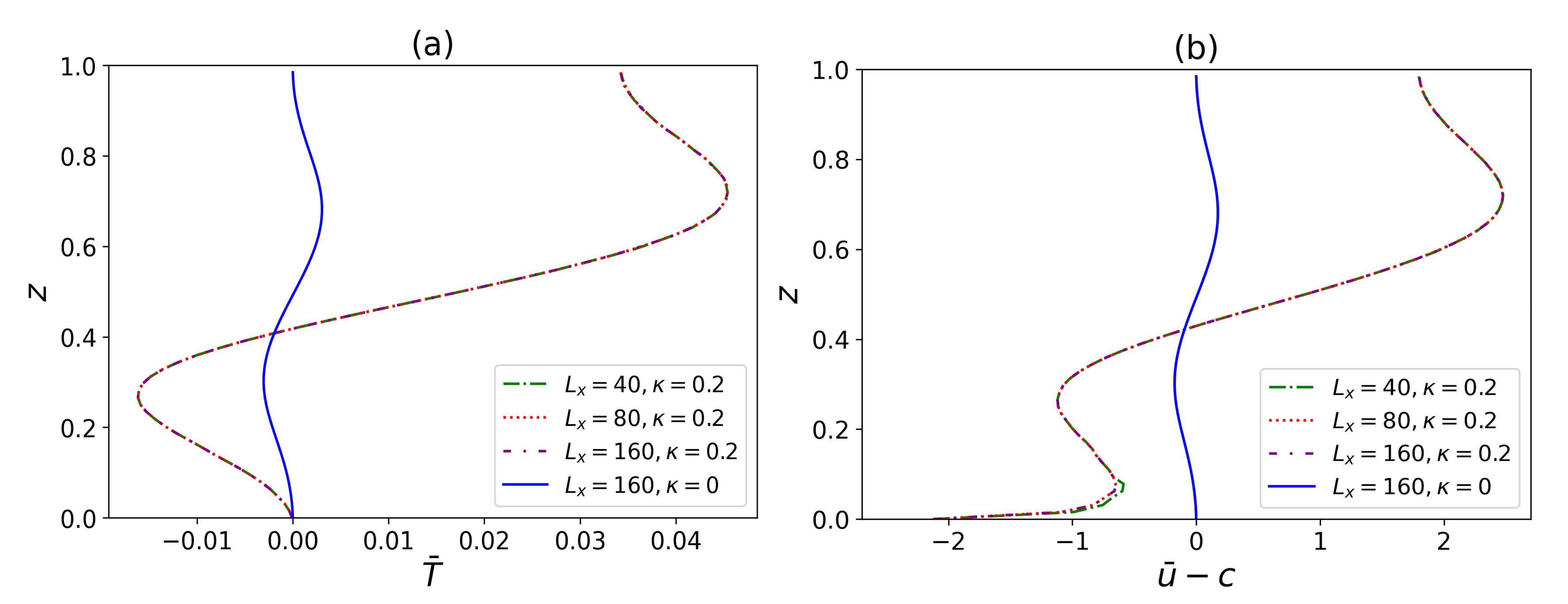}
\caption{\textcolor{black}{Horizontally averaged profiles of (a) the temperature $\bar{T}(z)$ and (b) the horizontal velocity $\bar{u}(z) - c$ in the comoving frame. Each panel includes the mean profiles from DNS of a single traveling pulse in domains of $L_x = 40$, $80$ and $160$ at $\kappa = 0.2$, with the corresponding spatial eigenvalues and drift velocity $c$ provided in Table \ref{tab:table1}. For comparison, the mean temperature and horizontal velocity profile of a stationary localized structure in an $L_x = 160$ domain with $\kappa = 0$ is also included (blue profiles). In both cases the averaged values are computed over the localized structure, defined by $|T(x, z)| \ge 10^{-4}$.}}
\label{fig:meanprofile}
\end{figure}

\begin{figure}[!htbp]
\includegraphics[width=1\linewidth]{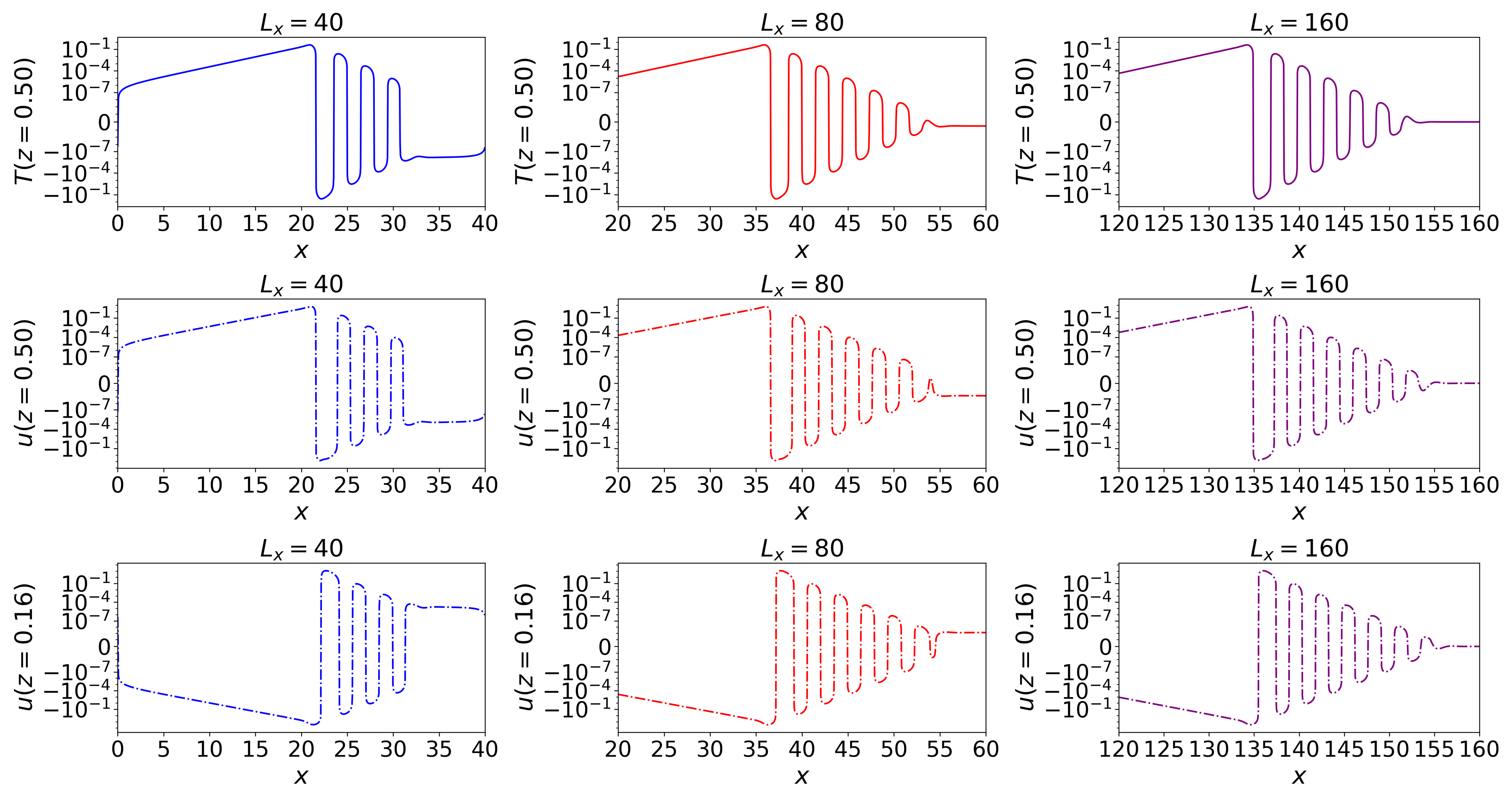}
\caption{\textcolor{black}{Temperature deviation profiles $T(x, z=0.5)$ (solid) and horizontal velocity deviation profiles $u(x, z=0.5)$ and $u(x, z=0.16)$ (dash-dot) of traveling single-pulse states at $\kappa = 0.2$, in $L_x = 40$, $L_x = 80$, and $L_x = 160$ domains. All data are shown with a symmetric logarithmic $y$ axis cutoff at $T(x, z) = \pm 10^{-10}$ and $u(x, z) = \pm 10^{-10}$. For the temperature field, profiles converge in domains with size $L_x \geq 80$. However, the velocity profiles exhibit noticeable differences between $L_x = 80$ and $L_x=160$ at both the midplane and $z=0.16$, indicating a different convergence rate compared with the temperature field.}}
\label{fig:profileatdifz}
\end{figure}

The dependence of the spatial eigenvalue predictions ($\lambda_2$ and $\lambda_3$) on the domain size arises entirely from changes in the drift speed $c$ as determined from DNS (Table~\ref{tab:table1}). As shown in Sec.~\ref{sec:section3}, the domain size has a significant impact on the drift speed of traveling localized structures: as $L_x$ increases, the drift speeds measured in DNS reflect better and better its behavior in an infinite domain, leading to a more accurate spatial eigenvalue prediction. \textcolor{black}{There are in fact two distinct contributions that set the speed $c$ in an inclined domain, the interaction between the tails of the pulse and its periodic images upslope and downslope, and the asymmetry in the buoyancy profile at the pulse location when $\kappa$ is nonzero. Figure~\ref{fig:meanprofile} illustrates this asymmetry in the mean temperature profile, $\bar{T}(z)$, and the horizontal velocity in the comoving frame, $\bar{u}(z) - c$. In sufficiently large domains the former almost vanishes (i.e.,~$\bar{T}(z)$ is almost antisymmetric), while the latter does not. The latter drives distinct corrections to the base flow (the mean flow correction $\bar{u}-c$ along the bottom boundary differs from that along the top boundary). Because of the asymmetry between the upslope and downslope tails when $\kappa>0$, changes in $L_x$ (and hence in $c$) affect $\beta_2$ and $\beta_3$ differently, although both depend on $L_x$ only weakly. Moreover, $c$ appears to have a stronger dependence on the domain size $L_x$ than either $\beta_2$ or $\beta_3$ (Table~\ref{tab:table1} and \ref{tab:table2}). Figure~\ref{fig:profileatdifz} shows why this is so. The figure shows that the tail profiles of $T(x,z=0.5)$ (and hence $\beta_2$ and $\beta_3$) have converged by $L_x=80$ but that both $u(x,z=0.5)$ and $u(x,z=0.16)$ (and hence $c$) continue to exhibit domain size dependence at this value of $L_x$. However, both $c$ and the associated change in $c$ remain small compared with the background mean flow $\boldsymbol{u}_{bs}=Ra\;\sin\;\phi (\frac{1}{2}-z)\boldsymbol{e}_x$ with $Ra=100$ as here, and so represent a small effect. The pointwise behavior shown in Fig.~\ref{fig:profileatdifz} thus differs substantially from that exhibited by the mean profiles $\bar{T}$ and $\bar{u}$ in Fig.~\ref{fig:meanprofile}. }

\begin{table}[!htbp]
\centering
\begin{tabular}{lllll} 
\hline\hline
Domain Size  & $L_x = 20$       & $L_x = 40$       & $L_x = 80$   & $L_x = 160$   \\ 
\hline
DNS fitted spatial growth rate $\beta_2$         & 0.63499049~ ~ ~~ & 0.63502656~ ~ ~~ & 0.63496979   & 0.63498433    \\
Spatial eigenvalue prediction of $\lambda_2$ & 0.5805~ ~ ~~     & 0.6058~ ~ ~~     & 0.6192~ ~ ~~ & 0.6260~ ~ ~~  \\
Drift velocity $c$ & 0.3398~ ~ ~~     & -0.2126~ ~ ~~     & -0.4888~ ~ ~~ & -0.6270~ ~ ~~  \\
\hline\hline
\end{tabular}
\caption{Spatial growth rate $\beta_2$ of the monotonic downslope tail fitted from DNS of the single-pulse localized structure at $\kappa = 0.2$ compared with the spatial eigenvalue prediction $\lambda_2$ using Eq.~\eqref{eq:spatial_eigenvalue}, together with drift velocities $c$ obtained from DNS that are used for the spatial eigenvalue prediction in each domain. 
}
\label{tab:table1}

\centering
\begin{tabular}{lllll} 
\hline\hline
Domain Size                      & $L_x = 20$             & $L_x = 40$         & $L_x = 80$       & $L_x = 160$       \\ 
\hline
DNS fitted spatial growth rate $\beta_3$     & -1.29325158~ ~ ~ ~ ~ ~ & -1.34512544~ ~ ~ ~ & -1.34617153~ ~ ~ & -1.34414370~ ~ ~  \\
Spatial eigenvalue prediction of~Re[$\lambda_3$] & -1.3254                & -1.3345            & -1.3402          & -1.3434           \\
DNS fitted wavenumber $k$     & 2.14466058             & 2.15905428         & 2.14943711       & 2.14943711        \\
Spatial eigenvalue prediction of~Im[$\lambda_3$] & 2.3244                 & 2.2340             & 2.1909           & 2.1698            \\
\hline\hline
\end{tabular}
\caption{Spatial growth rate $\beta_3$ and wave number $k$ of the upslope oscillatory tail fitted from DNS of the single-pulse localized structure at $\kappa = 0.2$ compared with the spatial eigenvalue prediction $\lambda_3$ from Eq.~\eqref{eq:spatial_eigenvalue}. Drift velocities $c$ used for spatial eigenvalue prediction are the same as shown in Table~\ref{tab:table1}.} 
\label{tab:table2}
\end{table}

\subsection{Transition in tail structure predicted by spatial eigenvalues}

\begin{figure}[!htbp]
\includegraphics[width=1\linewidth]{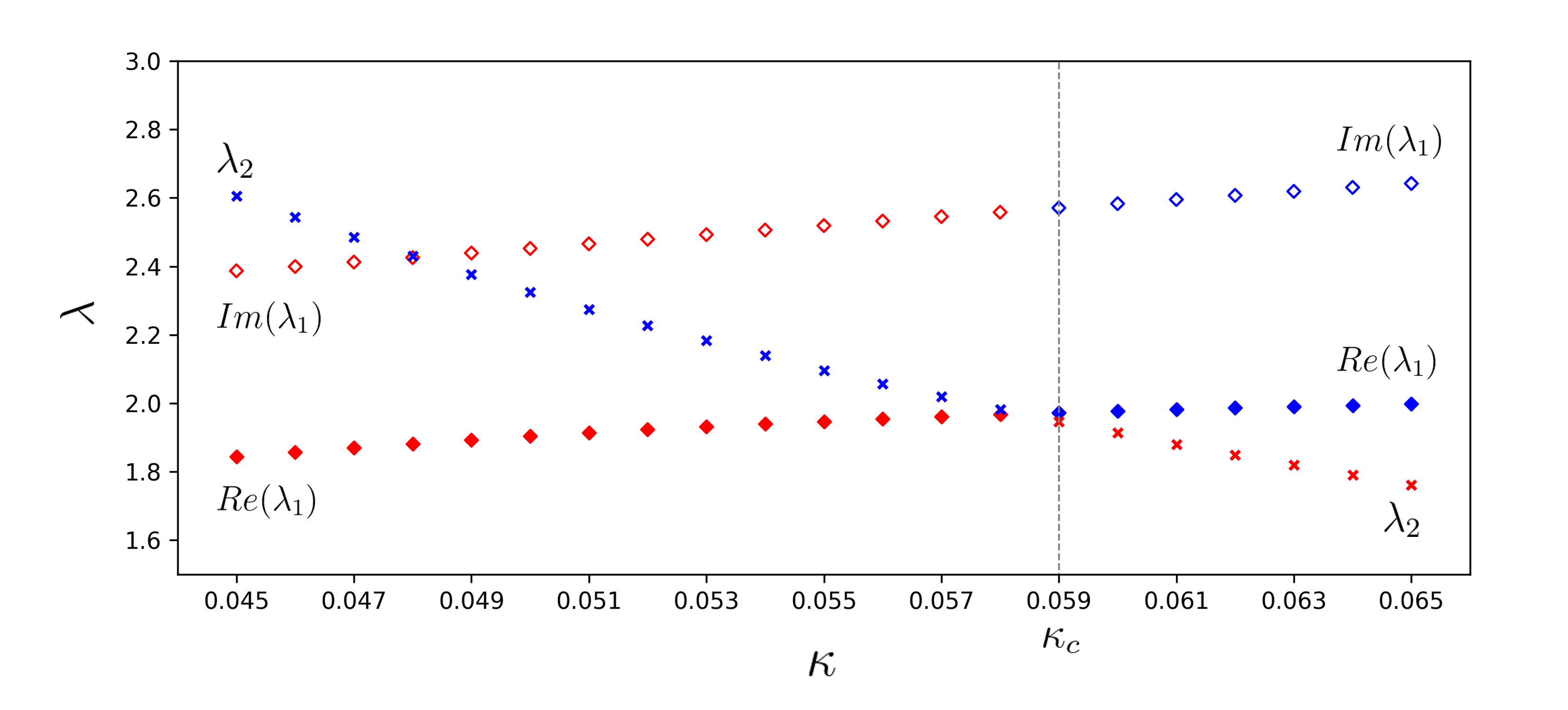}
\caption{Spatial eigenvalues $\lambda_1$ (diamond markers) and $\lambda_2$ (cross markers) associated with the downslope tail of a traveling single-pulse state as a function of $\kappa$. Red markers indicate the dominant spatial eigenvalues. When $\kappa < \kappa_c$, the tail is dominated by $\lambda_1$ and its complex conjugate $\lambda_1^*$, and is therefore oscillatory; when $\kappa > \kappa_c$, the tail is dominated by $\lambda_2$, and is monotonic. } 
\label{fig:spaeigpre}
\end{figure}

Figure~\ref{fig:spaeigfit}, discussed earlier, shows that the (leading) downslope tail of the traveling spatially localized structures can be oscillatory or monotonic depending on the value of $\kappa$. This property is determined by the competition between the complex eigenvalues $\lambda_1, \lambda_1^*$ and the real eigenvalue $\lambda_2$ shown in Fig.~\ref{fig:spaeigcomp}. To understand when the transition in tail structure occurs, we computed the spatial eigenvalues associated with the leading downslope tail in the regime of weakly broken symmetry (small $\kappa$). Figure~\ref{fig:spaeigpre} shows that as $\kappa$ increases, $\rm Re(\lambda_1)$ increases while $\lambda_2$ decreases.  At a critical value of $\kappa_c \approx 0.059$, there is a crossover where for $\kappa>\kappa_c$, $\lambda_2$ takes over the role of the dominant eigenvalue from $\lambda_1$, i.e., $\lambda_2$ has a real part closer to the origin than $\lambda_1$. This process and a similar picture can be found in the work of Knobloch and Yochelis \cite{knobloch2021stationary} on a multivariable reaction-diffusion system, where a similar exchange point plays a key role.

\begin{figure}[!ht]
\includegraphics[width=1\linewidth]{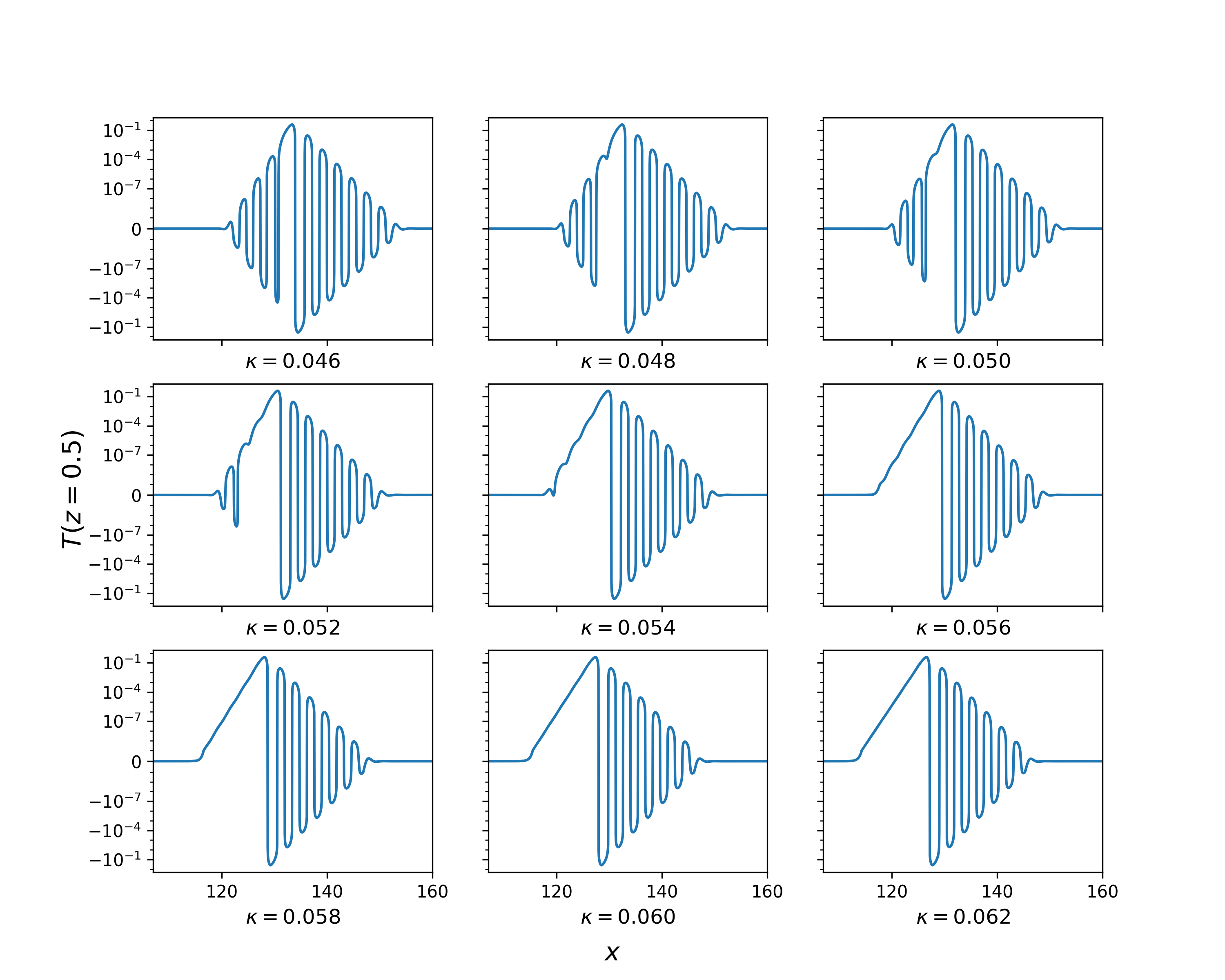}
\caption{Midplane temperature perturbation profiles $T(x,z=0.5)$ of traveling single-pulse states at different values of $\kappa$ in a wide domain with aspect ratio $L_x = 160$. All data is shown with a symmetric, logarithmic $y$ axis cut off at $T(x,z=0.5)=\pm 10^{-10}$. For all cases shown, the single-pulse structure travels downslope, i.e. in the $-\mathbf{e}_x$ direction. The leading (downslope) tails of the pulses are oscillatory when $\kappa$ is well below $\kappa_c = 0.059$ and gradually morph into a monotonic, exponentially decaying profile as $\kappa$ increases past $\kappa_c$ (through a range where the two forms are superposed). } 
\label{fig:spaeigtran}
\end{figure}

The spatial eigenvalues shown in Fig.~\ref{fig:spaeigpre} suggest that for a value of $\kappa$ below the critical value $\kappa_c$, the downslope tail of the traveling spatially localized structures is oscillatory. For $\kappa$ above the critical value $\kappa_c$, the downslope tail becomes monotonic. To confirm this transition in terms of the tail structure, we simulate traveling single-pulse states in a wide domain of aspect ratio $L_x = 160$ at different values of $\kappa$ near the critical exchange point $\kappa_c$. Figure~\ref{fig:spaeigtran} shows how the downslope oscillatory tail of a traveling single-pulse state in a $L_x = 160$ domain gradually deforms into a monotonic tail as $\kappa$ crosses the critical value $\kappa_c$. As shown in Fig.~\ref{fig:spaeigtran}, at $\kappa = 0.046$, both tails of the localized structure are oscillatory. As $\kappa$ increases, the downslope tail gradually becomes monotonic, while the upslope tail remains oscillatory, and the downslope tail becomes completely monotonic once $\kappa$ is greater than the critical value $\kappa_c\approx 0.059$. The transition is gradual rather than abrupt, because when the real parts of $\lambda_1$ and $\lambda_2$ are comparable, a superposition of the two associated exponential tails emerges as a result. In summary, these observations confirm the existence of a critical value $\kappa=\kappa_c$ distinguishing different downslope tail profiles and the corresponding prediction based on spatial eigenvalues.

\section{Long-time temporal dynamics of interacting localized structures}
\label{sec:section5}

Section~\ref{sec:section3} shows that at late times, traveling states at small $\kappa$ take the form of compact bound states (Fig.~\ref{fig:t2p_0.01}), while at large $\kappa$ pulses repel each other leading to an equispaced state (Fig.~\ref{fig:t2p_0.5}). The long-range interaction between spatially localized structures has been widely studied and is linked to the form of the exponential tails of these localized structures. Specifically, when the overlapping tails are both oscillatory (corresponding to complex spatial eigenvalues) then bound states exist generically, as established theoretically for numerous systems \cite{gorshkov1981interactions,aranson1990stable,vladimirov2002two,tlidi2003interaction,tlidi2008vegetarian,clerc2010interaction}. For instance, such multipulse bound states are found in the Swift-Hohenberg equation \cite{burke2009multipulse}. In contrast, in the case of monotonic tails (real spatial eigenvalues) localized structures typically repel one another, see e.g. \cite{ei1994equation,ohta2001pulse,berrios2020repulsive,eiermann2004bright,knobloch2021stationary}. In the model considered here, the upslope tail of the localized structures remains oscillatory (complex spatial eigenvalue) as $\kappa$ is increased, while the downslope tail undergoes a transition from oscillatory to monotonic form at the threshold $\kappa = \kappa_c \approx 0.059$. As a result, for $\kappa < \kappa_c$, the interaction between adjacent pulses is mediated by two oscillatory tails, which can potentially lock to one another, leading to bound states. By contrast, for $\kappa > \kappa_c$, the interaction between adjacent pulses is mediated by the overlap between a monotonic tail and an oscillatory tail, which typically leads to repulsion (but not in all cases, see also Sec.~\ref{sec:section6}). 

In this Section, we analyze further the long-time temporal dynamics of the repulsive interaction between traveling spatially localized structures by measuring the separation between pulses as a function of time at different values of $\kappa$. We extend selected DNS from Sec.~\ref{sec:section3} to investigate the long-time temporal dynamics of both bound and repelling traveling spatially localized structures, studying the time evolution of the inter-pulse separation in the vicinity of the threshold $\kappa=\kappa_c$.

\subsection{Bound states for $\kappa < \kappa_c$}

\begin{figure}[!htbp]
\includegraphics[width=0.9\linewidth]{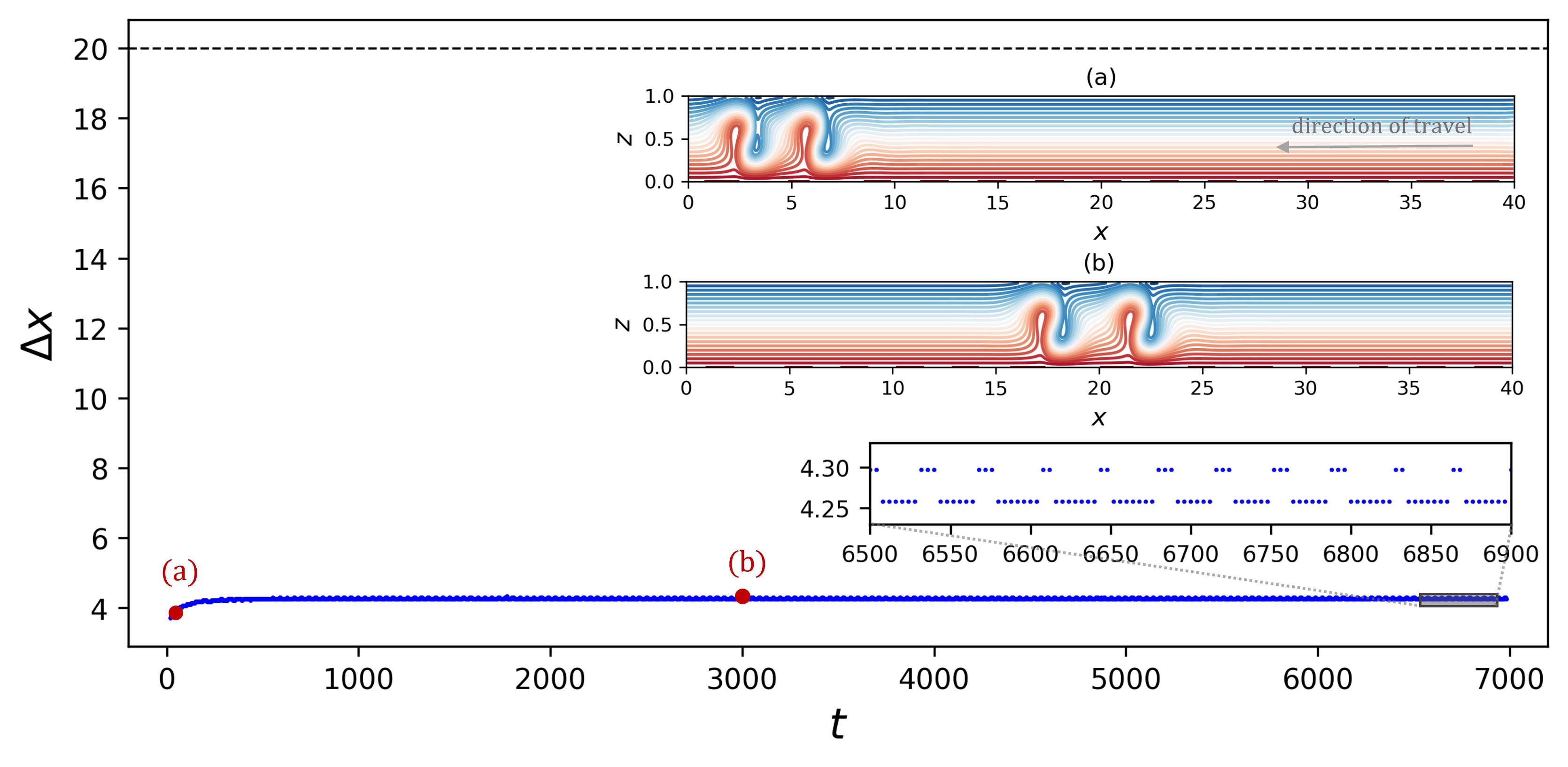}
\caption{Separation $\Delta x$ between adjacent pulses versus time for a traveling two-pulse structure in a $L_x = 40$ domain, at $\kappa = 0.04 < \kappa_c$, together with their total temperature profiles $T_{tot}(x, z)$ at (a) $t = 21$ and (b) $t = 2960$. The separation between the two pulses increases with time and saturates at a distance of about $\Delta x=4.25$. A zoom-in at the late time $t\in[6500, 6900]$ shows that the separation $\Delta x$ fluctuates within one unit of grid resolution. This final separation is much smaller than the largest possible separation $\Delta x = 20$ (marked by a dashed line) for two-pulse localized structures in a $L_x = 40$ domain. Profile (b) also shows that the two-pulse structure is far from an equispaced traveling state, suggesting the existence of the traveling bound state.}
\label{fig:sep004}
\end{figure}

\begin{figure}[h]
\includegraphics[width=1\linewidth]{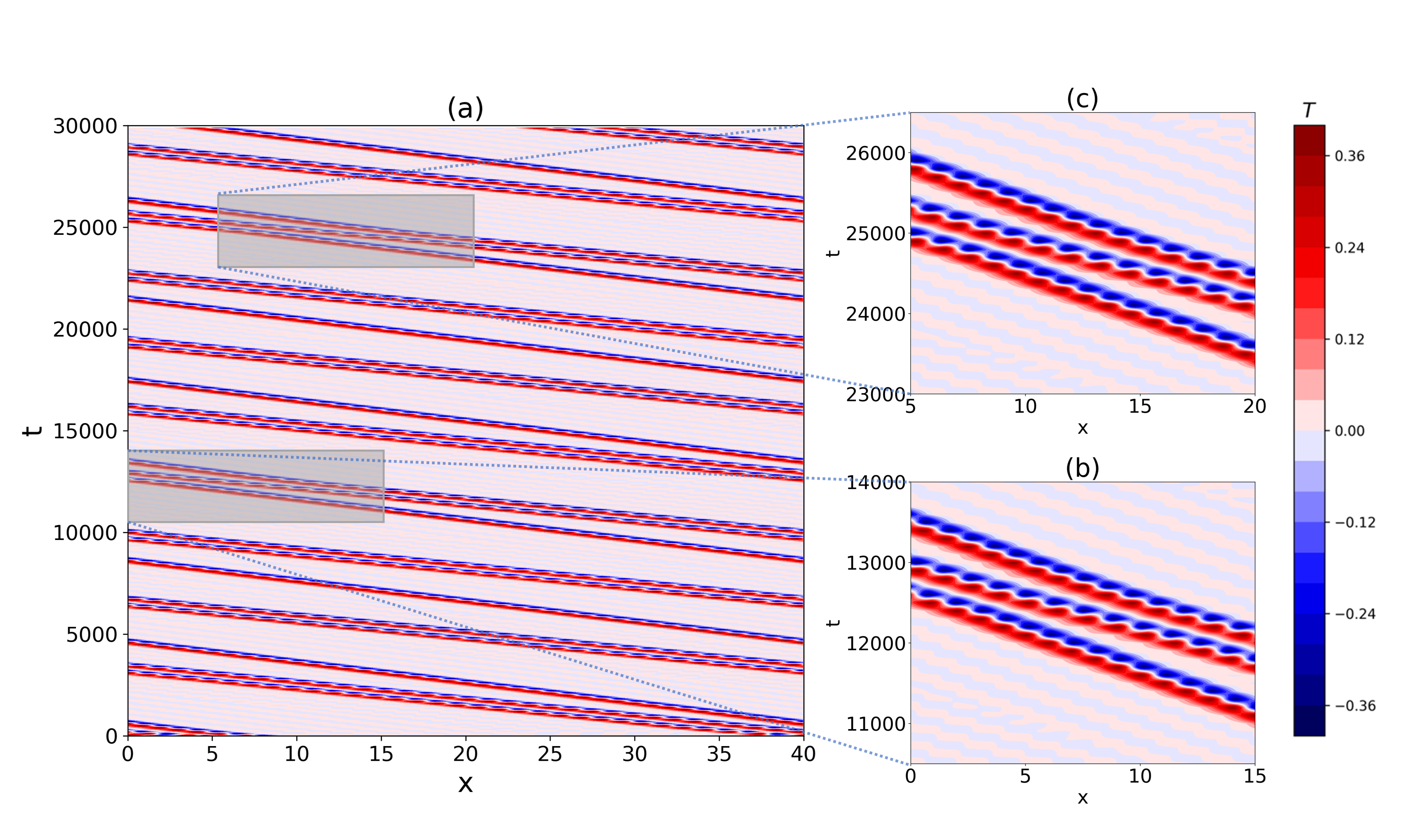}
\caption{Space-time plot, with time along the vertical axis and space along the horizontal axis, of a simulation initialized with a three-pulse spatially localized structure in a domain of aspect ratio $L_x = 40$ at $\kappa = 0.04 < \kappa_c$. Colors represent the midplane temperature deviation $T(x, z=0.5,t)$. Panel (a) shows the space-time plot from $t = 0$ to $t = 30000$. Panels (b) and (c) zoom-in at the first and second collision between the deviated single pulse and the two-pulse bound state. \textcolor{black}{The apparent modulation of the profile is a graphical artifact due to low output frequency; we reiterate that each pulse maintains its form while drifting, without any oscillations.}}
\label{fig:3p004}
\end{figure}

In Fig.~\ref{fig:t2p_0.01}, we have already seen that for weak symmetry breaking (at $\kappa=0.01$), traveling bound states of up to five pulses exist. Here, we first examine a traveling two-pulse structure in a $L_x = 40$ domain but with a slightly stronger (albeit still weak) symmetry breaking of $\kappa = 0.04$, which is near but still below the critical threshold $\kappa_c$. The separation between the pulses is measured over time as shown in Fig.~\ref{fig:sep004}. After a short initial transient, where the separation $\Delta x$ between the two pulses increases only slightly, the configuration quickly converges to $\Delta x \approx 4.25$, \textcolor{black}{with fluctuations within one unit of grid resolution, as shown in the zoom-in at late times in Fig.~\ref{fig:sep004}.}
This separation is significantly less than the maximum possible separation $\frac{1}{2} L_x = 20$, which indicates the presence of a true bound state, whose size is independent of $L_x$. The insets in Fig.~\ref{fig:sep004} showing the total temperature profiles of the localized structures at an early time (a) and at a later time (b) further illustrate that the two pulses in the structure remain bound, with only a modest increase in separation, relative to the domain size, as the initial condition from $\kappa=0$ adjusts to $\kappa=0.04$.

\begin{figure}[!htbp]
\includegraphics[width=0.7\linewidth]{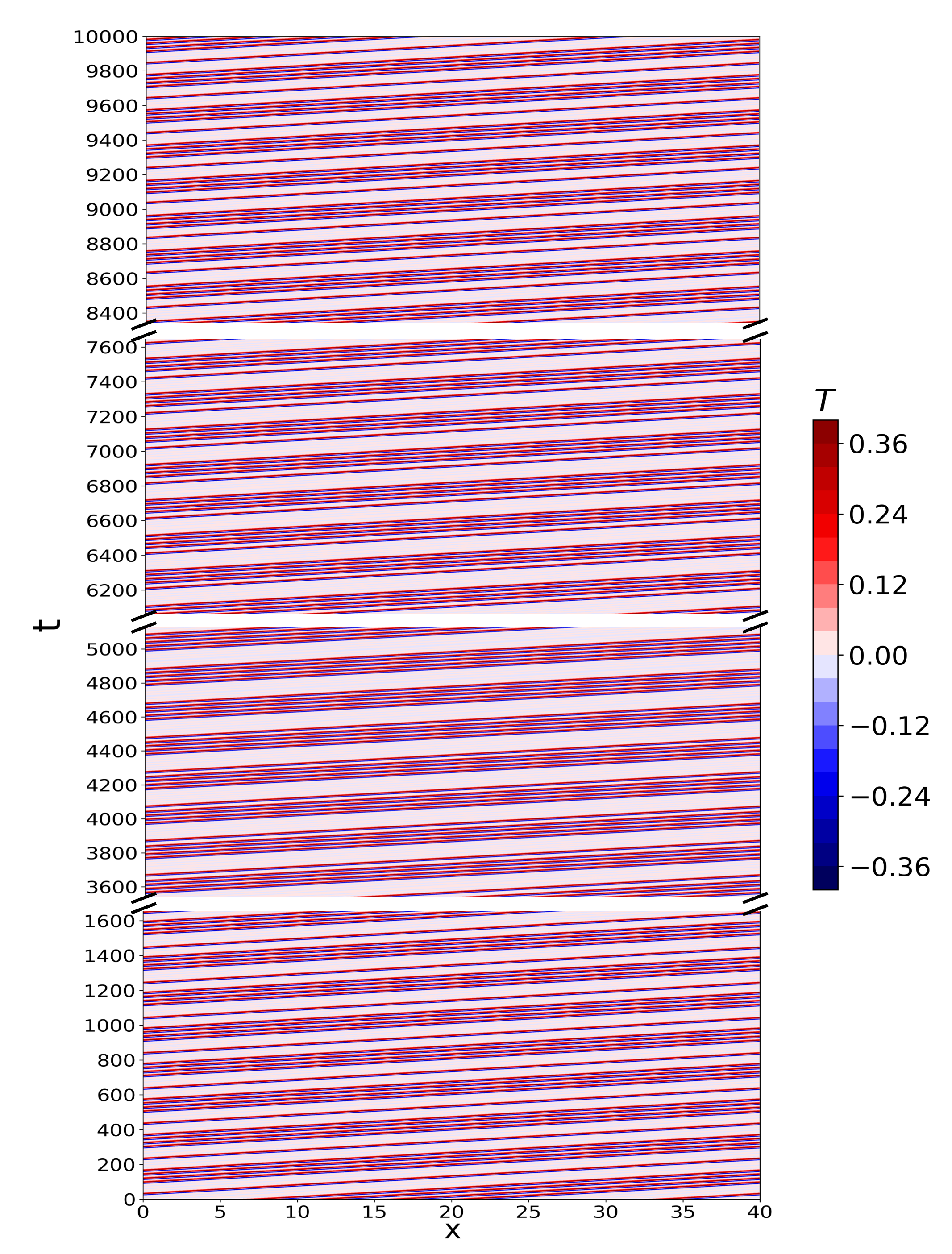}
\caption{Space-time plot of a traveling five-pulse structure in a domain of aspect ratio $L_x = 40$ at $\kappa = 0.04 < \kappa_c$. Colors represent the midplane temperature deviation $T(x, z=0.5,t)$, shown with time along the vertical axis and space along the horizontal axis.}
\label{fig:5p004}
\end{figure}

While the four-pulse traveling bound state is also observed to be stable in $L_x = 40$ domain at $\kappa = 0.04$, (i.e., somewhat below the critical value $\kappa_c$), the three-pulse and five-pulse structures are found to be unstable at the same parameter values as a result of the repulsive interaction. As shown in Fig.~\ref{fig:3p004}(a), the right-most, trailing pulse in a flow initialized with a three-pulse spatially localized structure in a domain of aspect ratio $L_x = 40$ at $\kappa = 0.04$ detaches from the other two pulses, drifting downslope more slowly than the remaining two-pulse bound state. As these structures continue to travel, the single pulse is eventually caught by the two-pulse bound state due to periodicity and a collision ensues. The first such collision occurs at $t \approx 12000$, a zoom-in of which is shown in Fig.~\ref{fig:3p004}(b). In this collision, the bound state breaks up and one of the constituent pulses binds with the single pulse, accompanied by the repulsion of another single pulse on the other side, and the process repeats. The ensuing evolution appears to be cyclic, with a similar event occurring around $t \approx 25000$ [see Fig.~\ref{fig:3p004}(c)], resembling the motion of Newton's cradle resulting from a perfectly elastic collision.
Similar behavior was also observed in a simulation initialized with a five-pulse spatially localized structure in a domain of aspect ratio $L_x = 40$ at $\kappa = 0.04$.
Figure~\ref{fig:5p004} shows related behavior resulting from an initial, unstable five-pulse structure at $\kappa=0.04$, where a similar Newton's cradle-like behavior ensues. In contrast with the three-pulse case, here the structures travel in the positive (upslope) direction, with the \textcolor{black}{leading} pulse gradually separating. However, in this scenario, this pulse is the rightmost pulse in the initial five-pulse structure. Due to periodicity, the detached single pulse eventually catches up with the four-pulse bound state at $t \approx 4500$ and the process repeats.

It is interesting to note that this type of instability of $n$-pulse states was only observed for $n=3$ and $n=5$, while two-pulse and four-pulse bound states were found to be stable over the long numerical integration time. This suggests that at this value of $\kappa$, the stability of $n$-pulse bound states depends on the parity of the integer $n$, with even $n$ leading to stable bound states and odd $n$ leading to instability via the detachment of the leading pulse. Further analysis will be needed to verify this hypothesis at larger values of $n$ and to provide a theoretical explanation. This is left for a future study.

%%%%%%%%%%%%%%%%%%%%%%%%%%%
\begin{figure}[!htbp]
\includegraphics[width=1\linewidth]{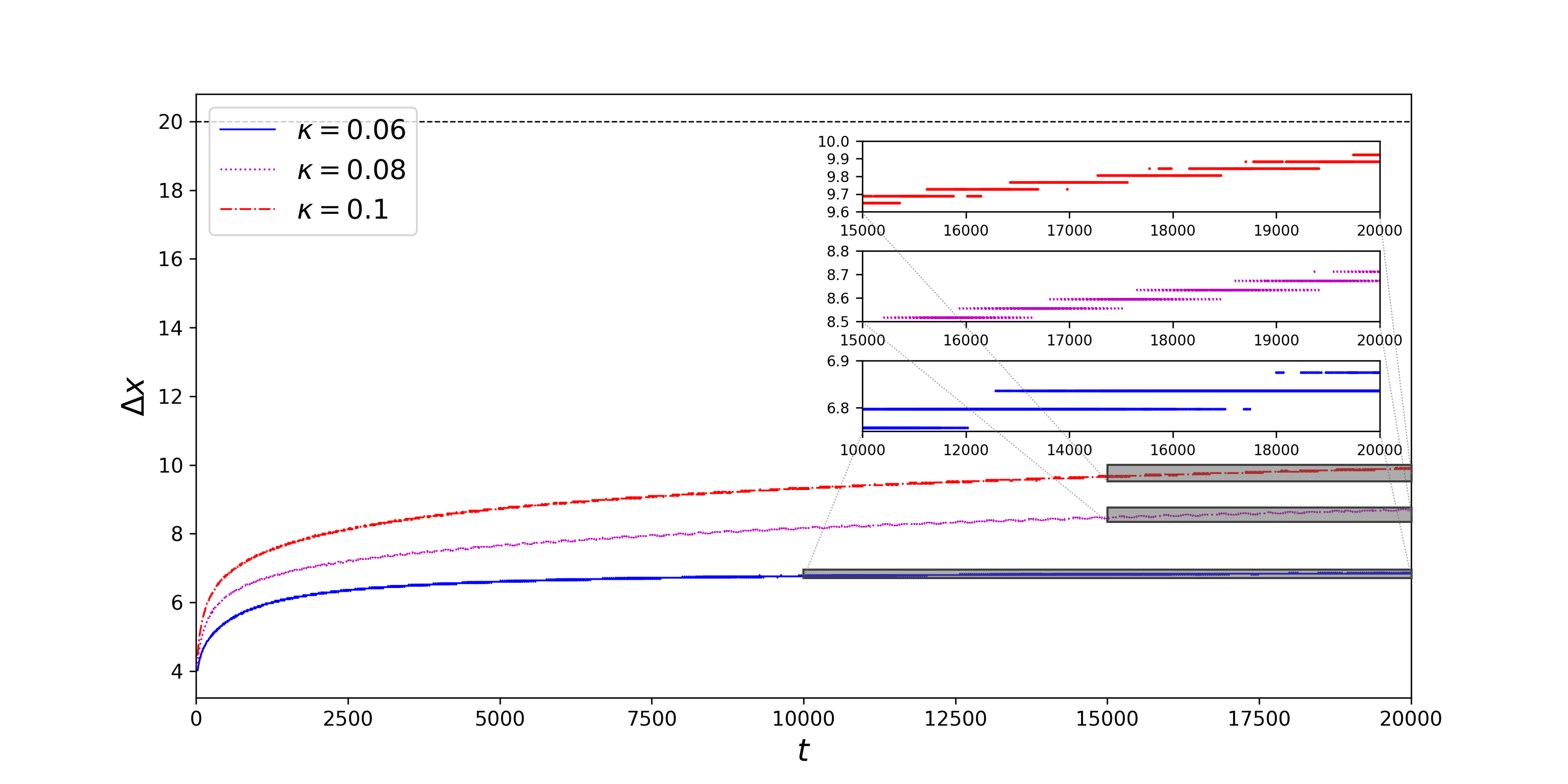}
\caption{Separation $\Delta x$ between pulses versus time for a traveling two-pulse structure in a $L_x = 40$ domain, at $\kappa = 0.06$, $\kappa = 0.08$ and $\kappa = 0.1$, with zoom-ins on $\Delta x(t)$ in the time range $t\in[10000, 20000]$ for $\kappa = 0.06$ and $t\in[15000, 20000]$ for $\kappa = 0.08$ and $\kappa = 0.1$. Note that these curves are discrete because the increment in separation is now equal to the grid resolution $\delta x=0.039$.}
\label{fig:40nonconv}
\end{figure}
%%%%%%%%%%%%%%%%%%%%%%%%%%%
%%%%%%%%%%%%%%%%%%%%%%%%%%%
\begin{figure*}[t]
\includegraphics[width=0.9\linewidth]{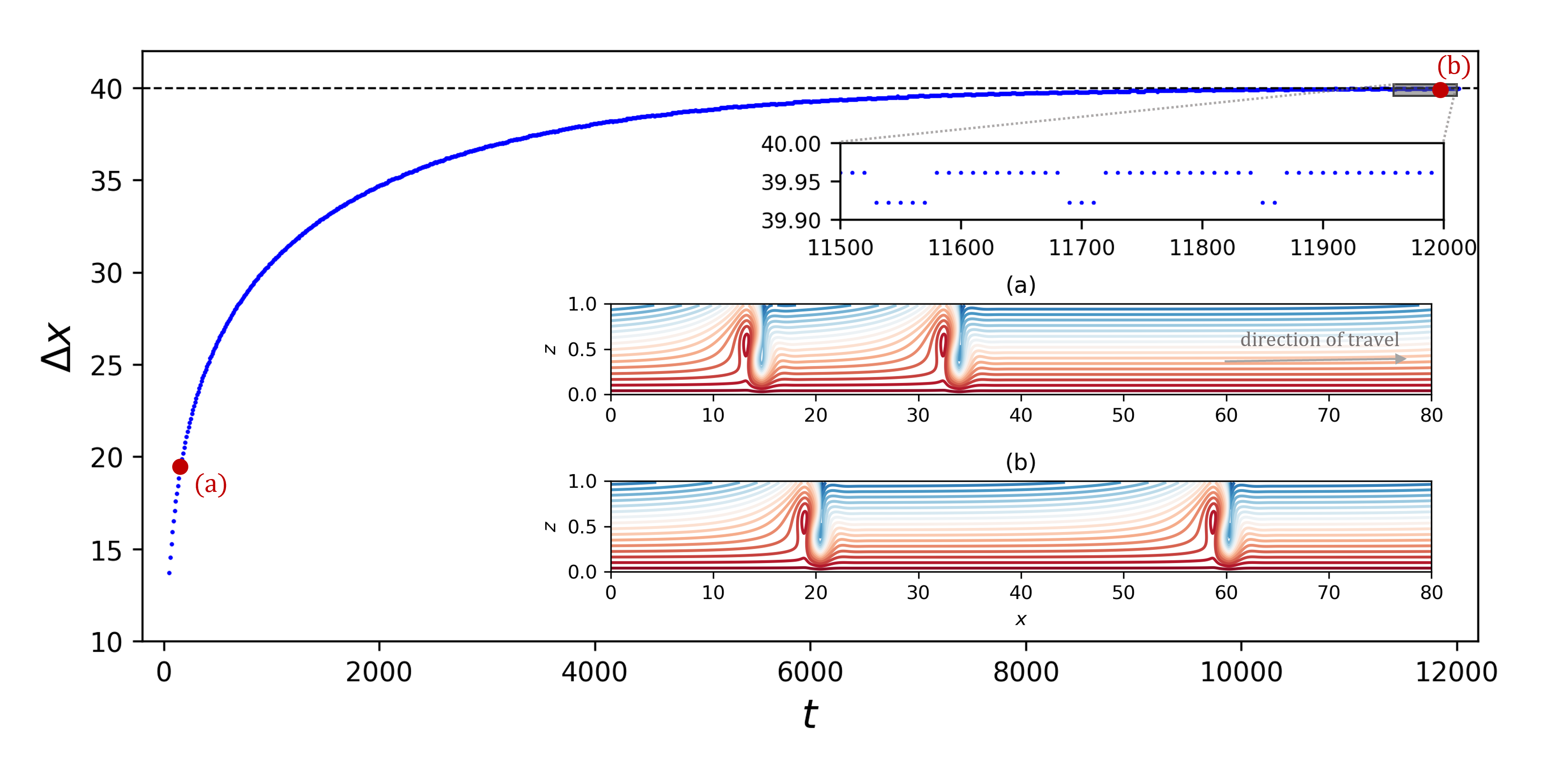}
\caption{Separation $\Delta x$ between pulses versus time for a traveling two-pulse structure in a $L_x = 80$ domain, at $\kappa = 0.95$, and the corresponding total temperature profiles $T_{tot}(x,z)$ at (a) $t = 160$ and (b) $t = 11980$. The separation between the two pulses increases with time and converges to the largest possible separation, $\Delta x = 40$ for two pulses in a $L_x = 80$ domain. A zoom-in at the late time $t\in[11500, 12000]$ shows that the separation $\Delta x$ fluctuates within one unit of grid resolution. As shown in profile (b), the traveling two-pulse structure is now equispaced in the domain.}
\label{fig:sep095}
\end{figure*}
%%%%%%%%%%%%%%%%%%%%%%%%%%%

\begin{figure}[h]
\includegraphics[width=1\linewidth]{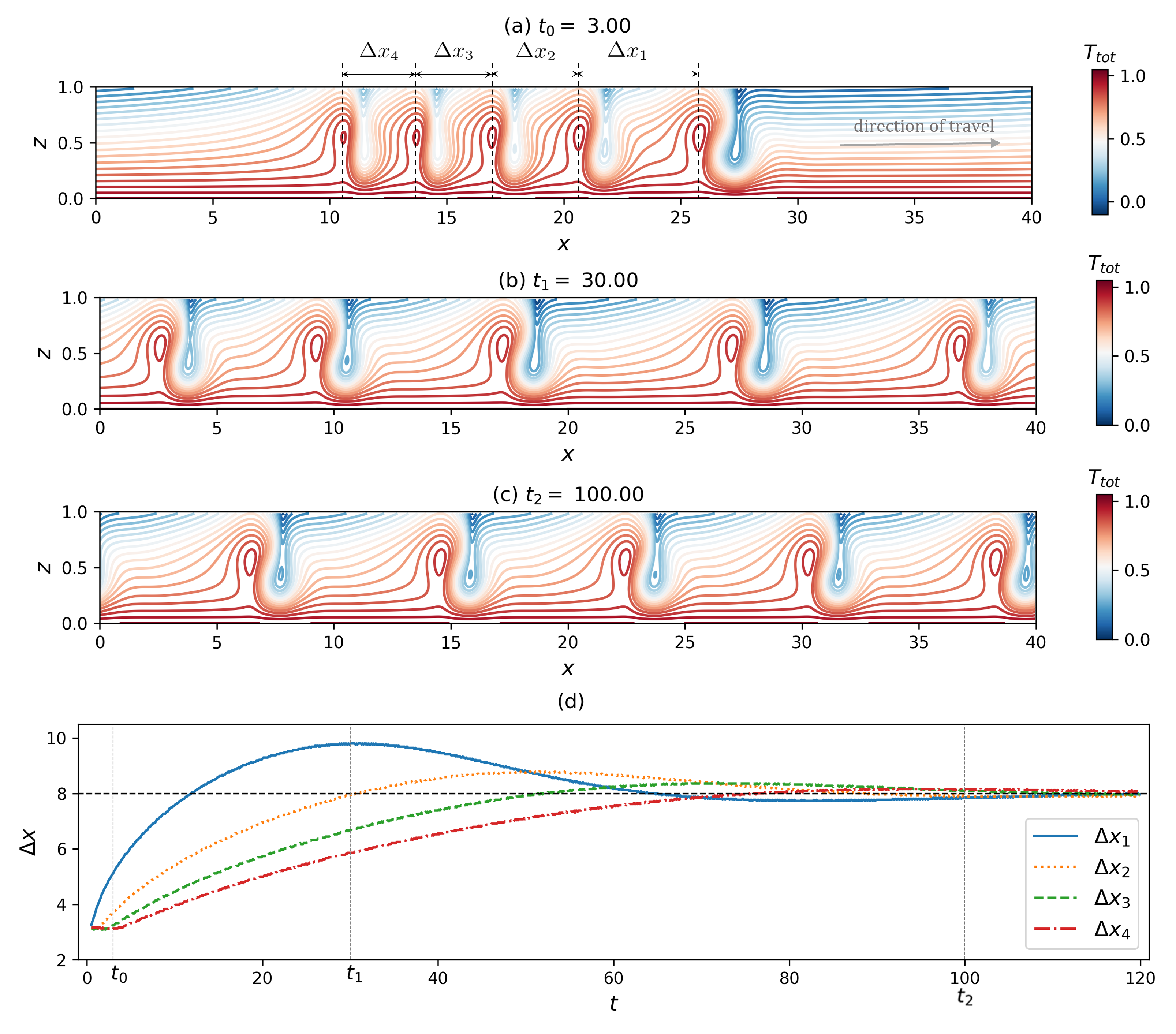}
\caption{Total temperature profiles $T_{tot}(x,z)$ of traveling spatially localized structures at time (a) $t_0 =3 $, (b) $t_1 = 30$, and (c) $t_2=100$ for a right-traveling five-pulse structure in a $L_x = 40$ domain with $\kappa = 0.9$. Separations between adjacent pulses are labeled $\Delta x_i$, $i\in\lbrace1,2,3,4\rbrace$, as indicated in panel (a), and their time dependence is shown in panel (d). The largest possible separation between five pulses in the $L_x = 40$ domain, $\Delta x = 8$, is marked out in the black dashed line in panel (d). The leading pulse experiences the strongest repulsion over time, resulting in a stronger overshoot of the separation $\Delta x_1$ and faster convergence to the final equispaced state.
} 
\label{fig:5pulserep}
\end{figure}

\subsection{Repulsive interaction for $\kappa > \kappa_c$} 

In Fig.~\ref{fig:t2p_0.5}, we demonstrated that pulses with a monotonic tail repel each other leading to equispaced states in finite domains (here $L_x=40$) for strong symmetry breaking (here $\kappa = 0.5$) and the coexistence of trains of between two and five pulses at identical parameter values. Similar equispaced traveling states were also observed in most cases where $\kappa > 0.5$, except when the traveling multi-pulse structures proved unstable, as in the case of the five-pulse structure at $\kappa = 1$. 

However, just as the drift speed depends on $\kappa$ and $L_x$ as discussed in Sec.~\ref{sec:section3}, the interactions between pulses can become exceedingly weak, particularly at small $\kappa$ or in larger domains, requiring very long simulations in order to determine unambiguously whether the system converges to a bound state with a small distance between pulses, which is intrinsically determined and independent of the domain size, or an equispaced state where all pulses are at the maximum distance from each other for the given domain size. In our simulations of traveling two-pulse localized structures in a $L_x = 40$ domain, we confirmed that for $\kappa > 0.2$ the pulses consistently repel each other to the maximum possible separation of $\Delta x = 20$ (not shown). For simulations with $\kappa \in [\kappa_c, 0.2]$, no bound states at finite separations less than the largest separation ($\Delta x = 20$ in this case) have been observed although it remains numerically challenging to determine unambiguously whether the system converges to an equispaced configuration or not due to the very long time scales involved. Figure~\ref{fig:40nonconv} shows the inter-pulse separation in a traveling two-pulse state versus time at $\kappa = 0.06, 0.08, 0.1$, respectively, with insets showing zoom-ins into the time interval $t\in[10000, 20000]$ for $\kappa = 0.06$ and $t\in[15000, 20000]$ for $\kappa = 0.08$ and $\kappa = 0.1$. 
Specifically, the separation between pulses in a traveling two-pulse spatially localized structure at $\kappa = 0.06$ only increases by four units of grid resolution in 10000 diffusive time units. \textcolor{black}{This is notably different from the behavior of the converged separation curve, with fluctuations of only one unit of grid resolution, shown in Fig.~\ref{fig:sep004}, indicating that the traveling two-pulse structure has yet to reach its final separation at this value of $\kappa$.} \textcolor{black}{
The slow dynamics shown in Fig.~\ref{fig:40nonconv} prevent us from fully determining whether the final state will be a bound state or an equispaced state using DNS, despite a simulation time of $t=20,000$. In Section \ref{sec:section6} we develop a reduced-order model that can be used to predict the final state in such parameter regimes.
}

In addition to the $L_x = 40$ domain, we also examined traveling two-pulse localized structures in an $L_x = 80$ domain and numerically confirmed the formation of equispaced traveling states at $\kappa = 0.95$. The separation between the two pulses is shown in Fig.~\ref{fig:sep095}. As the spatially localized structure travels, the constituent pulses repel each other and eventually reach a final state where they are equally spaced at half the domain size. The temperature deviation profiles at an early time (a) and a later time (b) clearly illustrate this change in separation over time, which contrasts sharply with the bound state observed at smaller $\kappa$ values shown in Fig.~\ref{fig:sep004}.

While repulsion also occurs in traveling multi-pulse structures with more than two pulses, the spreading process can involve more complex dynamics than that of the two-pulse localized structure. Figure~\ref{fig:5pulserep} illustrates the spreading dynamics resulting from the repulsive interaction. Panels (a), (b) and (c) of Fig.~\ref{fig:5pulserep} show snapshots of the full temperature field at different stages of the process. At very early times (panel (a)), the right-most pulse (which is also the leading pulse in this case since the structure travels to the right, i.e. upslope), is propelled to the right the fastest. Subsequently, the second pulse from the right detaches and moves to the right faster than the pulses further to the left (panel (b)). At late times (panel (c)), the system reaches an equispaced configuration where all pulses have spread out evenly. 

In Fig.~\ref{fig:5pulserep}(d), this evolution is quantified in terms of the separation between adjacent pulses over time. One observes that the separations systematically overshoot the final equidistant spacing, $\Delta x = 8$ here, with the rightmost pulse reaching the maximum separation from its adjacent pulse first, followed by the second from the right, etc. As $t$ increases to late times (i.e., $t \gtrsim 100$), the separation curves saturate at $\Delta x = 8$ while undergoing damped, small-amplitude oscillations, a nontrivial feature given the absence of inertia in the system.

\subsection{Collision between traveling bound states}

\begin{figure}[h]
\includegraphics[width=1\linewidth]{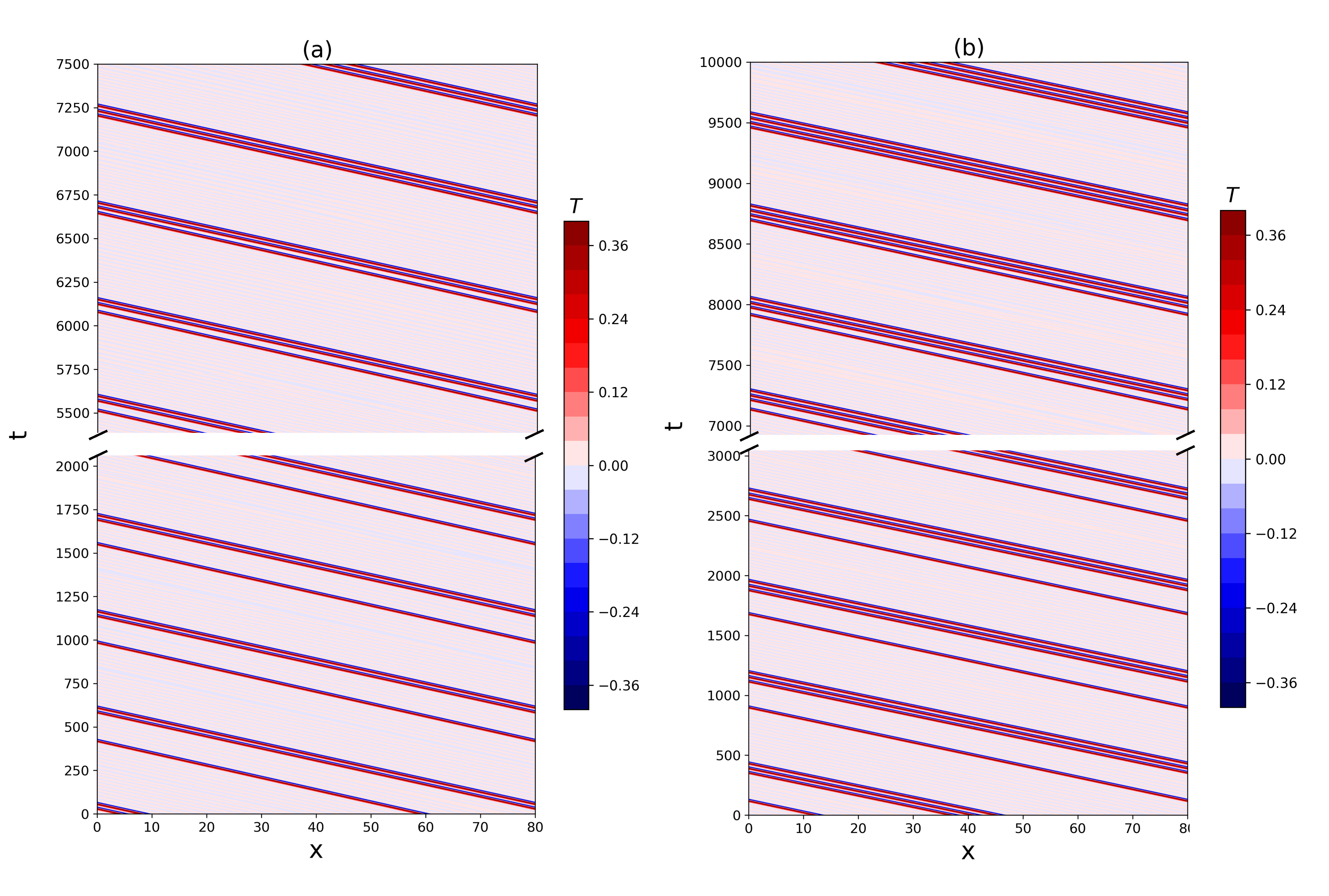}
\caption{Space-time plots of the midplane temperature deviation $T(x, z=0.5,t)$ displaying collisions between a single pulse and bound states consisting of two (panel (a)) or three pulses (panel (b)) at $\kappa=0.03$. The collisions are conservative, producing new bound states consisting of three and four pulses, respectively, without the creation or annihilation of additional pulses. 
} 
\label{fig:collisions2}
\end{figure}

The discovery of cyclic, Newton's cradle-like behavior of three-pulse and five-pulse traveling localized structures at $\kappa = 0.04 <\kappa_c$ motivates further investigation of collisions between traveling convectons in bound states. We utilized the drift velocity difference between traveling bound states with different numbers of pulses to set up chasing collision scenarios in DNS by concatenating two periodic domains in the $x$ direction initialized with different $n$-pulse bound states.

Figure~\ref{fig:collisions2} shows collisions between a single pulse and bound states consisting of two (panel (a)) or three pulses (panel (b)) at $\kappa=0.03$ in a combined domain of $L_x = 80$, formed by concatenating these two traveling localized structures within $L_x = 40$ domains. The collisions conserve the number of pulses, but are inelastic, producing simple combined bound states consisting of three and four pulses, respectively, without the creation or annihilation of pulses. This phenomenology differs from collisions reported previously in binary fluid convection \cite{mercader2013travelling}, which are significantly impacted by fluid inertia, and in the bistable Swift-Hohenberg equation \cite{houghton2011swift,raja2023collisions}, proposed as an order parameter description of spatially localized structures in binary fluid convection. In both cases, more complex dynamics ensue upon collision, including annihilation and creation of new pulses. Collisions of localized modulated traveling waves in binary fluid convection also lead to distinct behaviors \cite{iima2005collision,taraut2012collisions,watanabe2012spontaneous}, such as the gradual growth of the traveling wave envelope in the aftermath of the collision, similar to depinning dynamics \cite{ma2012depinning}, the annihilation of certain localized traveling waves, or the generation of new waves, none of which are observed here.

\textcolor{black}{In addition to collisions of bound states, we find unstable structures in the combined domain when concatenating structures which are individually stable at sufficiently large $\kappa$. For example, at $\kappa = 0.5$, when concatenating a two-pulse traveling structure with a single-pulse structure in two $L_x = 40$ domains, both individually stable in this domain, the resulting state in a domain of size $L_x = 80$ spontaneously decays to the base state solution immediately after initialization, in a manner identical to Case 2 reported in Sec.~\ref{sec:section3} C. This observation suggests a potential long-range interaction between localized structures (possibly mediated by incompressibility), though this requires further investigation.} Another possible explanation is that the fold of the three-pulse state is located at a larger Rayleigh number than that of the one- and two-pulse states, but numerical continuation techniques will be needed to assess this hypothesis.

%%%%%%%%%%%%%%%%%%%%%%%%%%%%%%
\section{Reduced-order modeling}\label{sec:section6}

Long-range interactions between localized structures are mediated by their exponentially growing or decaying oscillatory tails. Such interactions via exponential tails are found in a wide range of problems and have been studied extensively, see e.g. \cite{coullet1987nature,kawahara1988pulse,elphick1991interacting,ei1994equation,balmforth1994chaotic,kozyreff2011cavity,nishiura2022traveling}. Recently, in \cite{raja2023collisions}, a simple reduced-order model was introduced to quantitatively describe these interactions. \textcolor{black}{This reduced-order model was shown to reproduce a wide range of collisions with quantitative accuracy for asymmetric spatially localized structures within the cubic-quintic Swift-Hohenberg equation (SH35) \citep{raja2023collisions}.  Here, we adapt this model to the problem at hand. This reduced-order model describes the intrinsic traveling behavior of each localized structure based on their phase speed as obtained from DNS in Sec.~\ref{sec:section3}. Superposed on this motion are effects of the pulse-pulse interaction described in a particle-like manner, modeled using the tail structures of the interacting pulses informed by the spatial eigenvalues $\lambda_2$ and $\lambda_3$ obtained in Sec.~\ref{sec:section4}. }

As in the DNS described above, we consider two localized structures whose proximate, large-amplitude maxima in $T(x,z=0.5,t)$ [defined as those with $T(x,z=0.5,t)>0.3$] are located at positions $x_1$ and $x_2$ in $[0,L_x]$; the distance between the two proximate, finite-amplitude maxima is denoted by $\Delta x\equiv |x_2-x_1|$. Each localized structure propagates at a given speed, denoted by $c_1$ and 
$c_2$, in the absence of interactions, and interactions modify their propagation. The dominant spatial eigenvalue with positive real part is denoted by $\Lambda\equiv \Lambda_r + i \Lambda_i$, with $\Lambda = \lambda_1$ for $\kappa<\kappa_c$ and $\Lambda=\lambda_2$ for $\kappa>\kappa_c$ while $\lambda_3=\lambda_{3r} + i \lambda_{3i}$ remains the dominant eigenvalue with negative real part for all $\kappa$. We take into account the finite domain size $L_x$ and periodic boundary conditions to obtain the following reduced model equations, reminiscent of overdamped particle dynamics, but with a nontrivial interaction force:
 \begin{subequations}
 \label{eq:reduced_model}     
 \begin{align}
 \frac{\mathrm{d}x_1}{\mathrm{d}t} =& c_1 + g_1 \cos(\Lambda_i\Delta x + \varphi)e^{-\Lambda_r \Delta x } + g_2 \cos(\lambda_{3i}(L_x-\Delta x) + \varphi)e^{\lambda_{3r} (L_x-\Delta x)}, \label{eq:red_a}\\
 \frac{\mathrm{d}x_2}{\mathrm{d}t} =&  c_2 + g_1 \cos(\Lambda_i(L_x-\Delta x) + \varphi)e^{-\Lambda_r (L_x-\Delta x) } + g_2 \cos(\lambda_{3i}\Delta x + \varphi)e^{\lambda_{3r} \Delta x}. \label{eq:red_b}
 \end{align}
 \end{subequations}
Here $g_1$ and $g_2$ are unknown interaction amplitudes and $\varphi$ is an unknown interaction phase. We stress that the $g_1$ in Eqs.~(\ref{eq:red_a}) and (\ref{eq:red_b}) are identical since they both describe the interaction with a downslope tail, while $g_2$ is likewise identical between 
Eqs.~(\ref{eq:red_a}) and (\ref{eq:red_b}). These three unknown parameters, $g_1,g_2$ and $\varphi$, are determined by least-square fitting of $x_1,x_2$ to DNS data, see \cite{raja2023collisions}. In principle, there could be distinct phases $\varphi_1,\varphi_2$ when two oscillatory tails are present, but it is found empirically that one phase is sufficient to accurately reproduce DNS results. The model can be generalized to more than two interacting pulses, but we restrict attention to two interacting localized states only. We compare DNS and reduced model results in two cases: (1) a repulsive interaction where one tail is monotonic and the other oscillatory tail, and (2) a chasing collision between a single pulse and a two-pulse bound state where both tails are oscillatory.

\subsection{Repulsion for $\kappa>\kappa_c$}
\begin{figure}
    \centering
    \hspace{0.5cm} (a) \hspace{6.5cm} (b) \hspace{4cm}\\
    \includegraphics[width=0.4\linewidth]{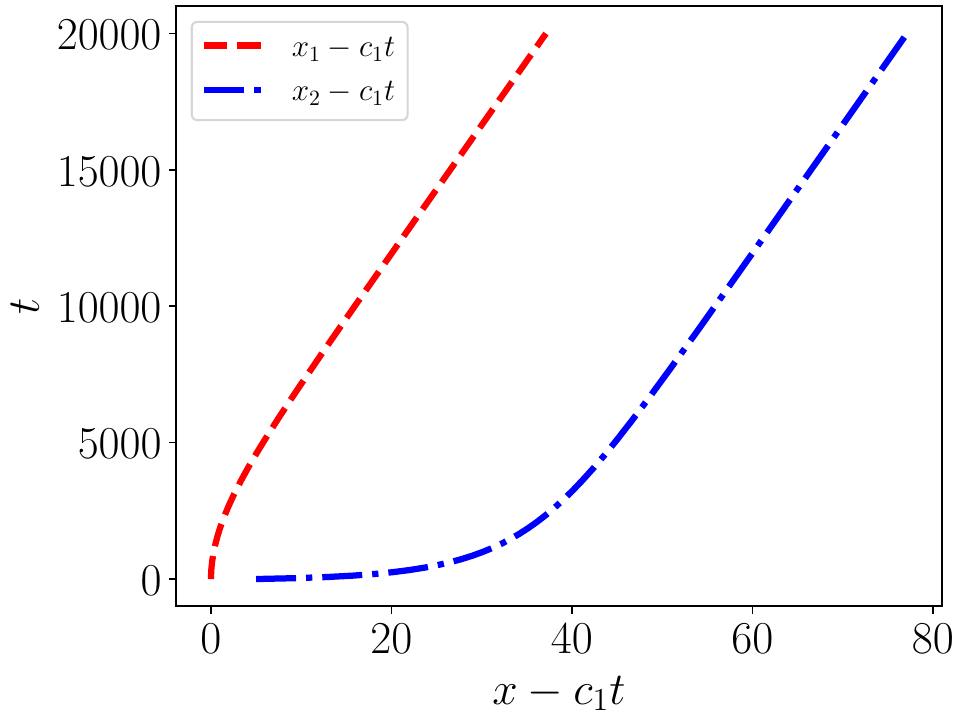}
    \includegraphics[width=0.4\linewidth]{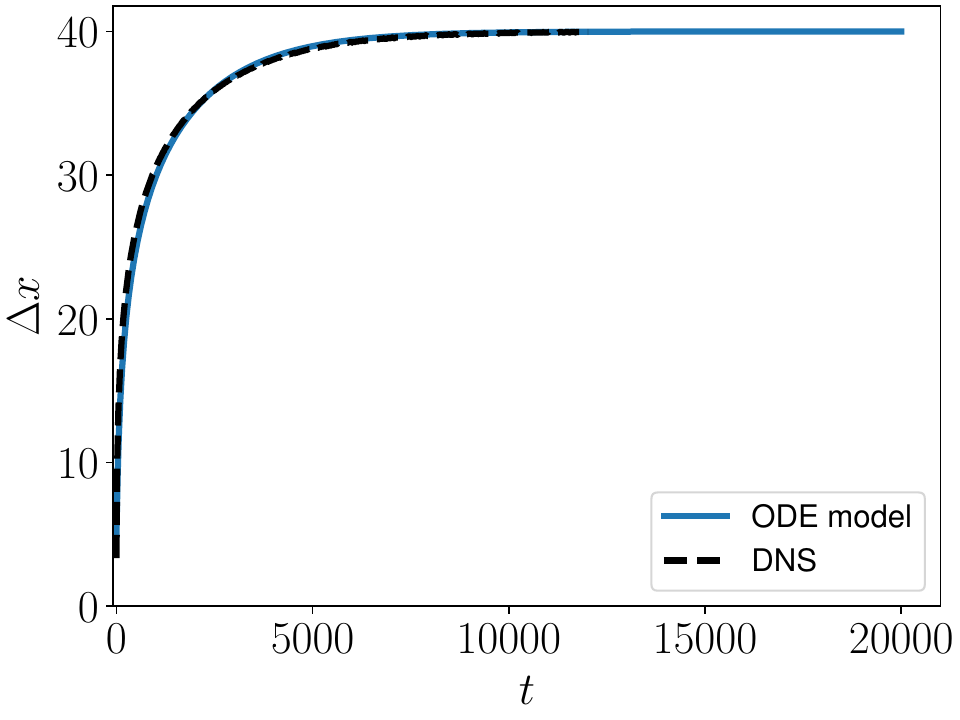}
    \caption{(a) Trajectories in the comoving frame of two identical interacting pulses, which in the absence of interactions propagate at velocity $c_1$, in a domain of size $L_x=80$ at $\kappa=0.95$ with spatial eigenvalues $\Lambda= 0.1274$, $\lambda_3=-1.35+ 2.56 i$, computed from the reduced model in Eq.~(\ref{eq:reduced_model}) with parameters $g_2=-g_1=0.347$, $\varphi=0$ estimated by fitting the trajectories to the corresponding DNS data. (b) The separation $\Delta x$ as a function of time for the two repelling pulses from DNS [see Fig.~\ref{fig:sep095}], and from the reduced model. The fact that the two curves overlap indicates that the reduced model accurately captures the dynamics observed in the DNS.}
\label{fig:ode_dns_repulsion}
\end{figure}

As in the DNS results shown in Fig.~\ref{fig:sep095}, we consider an initial two-pulse state composed of two identical single-pulse structures at $\kappa=0.95$ [where $\Lambda=\lambda_2\in \mathbb{R}_+$, i.e., $\Lambda_i=0$ in Eq.~(\ref{eq:reduced_model})] in a domain of size $L_x=80$. In isolation, each pulse travels with the same velocity $c_1=c_2$. In this case, the only equilibrium of Eq.~(\ref{eq:reduced_model}) is $\Delta x=L_x/2$. Indeed, this is confirmed by integrating the ODEs using a fourth-order Runge-Kutta time-stepping scheme, starting from the same initial positions of the pulses as in the DNS. Figure~\ref{fig:ode_dns_repulsion}(a) shows trajectories in the comoving frame of two interacting pulses at $\kappa=0.95$ in a domain of size $L_x=80$ from the reduced model in Eq.~(\ref{eq:reduced_model}) with parameters estimated by fitting the trajectories to the corresponding DNS data. The right-most pulse is propelled to the right faster initially, as seen in all repulsive interactions in DNS; cf. Fig.~\ref{fig:5pulserep}. Figure~\ref{fig:ode_dns_repulsion}(b) shows a comparison, for two interacting pulses at $\kappa=0.95$ in an $L_x=80$ domain, between the separations $\Delta x(t)$ obtained from DNS (cf.~Fig.~\ref{fig:sep095}) and the model result. The two curves agree well and we conclude that the particle-like dynamics of Eq.~(\ref{eq:reduced_model}) correctly describe the observed repulsion between localized pulses.

\begin{figure}
    \centering
    \includegraphics[width=\linewidth]{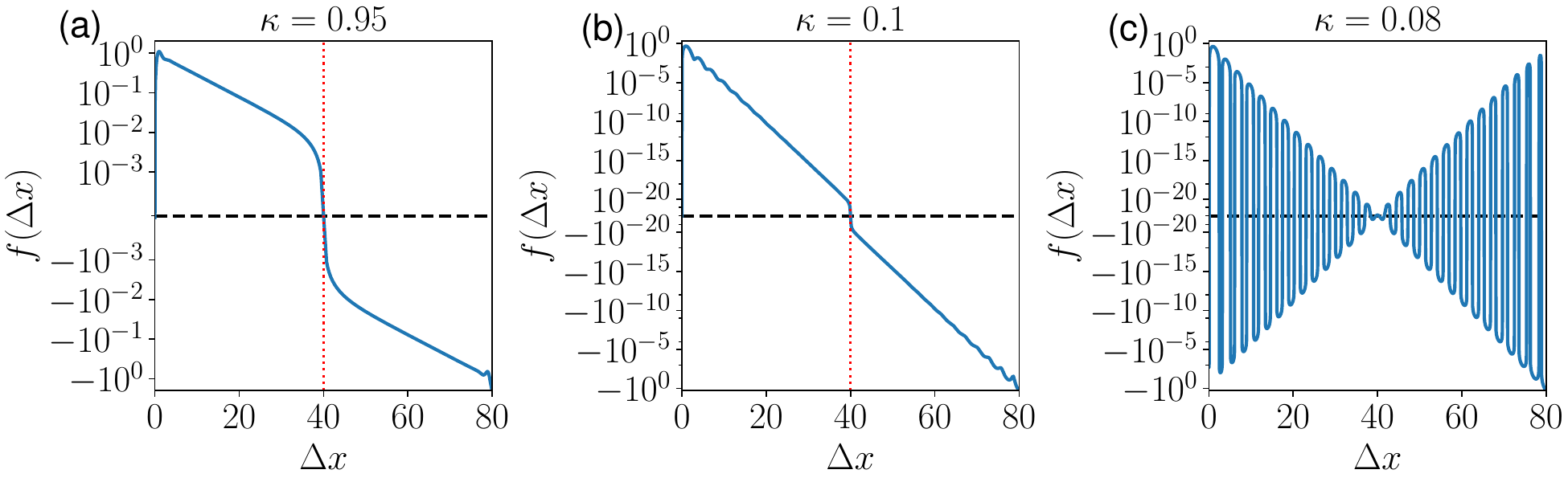}    
    \caption{Residual $f(\Delta x)$ defined in Eq.~(\ref{eq:dDelta_dt}) versus $\Delta x$ in a domain of size $L_x=80$ with spatial eigenvalues $\Lambda=0.1274$, $\lambda_3 = -1.35
+2.56i$ for $\kappa=0.95$ [panel (a)], $\Lambda= 1.176$, $\lambda_3=-1.2848+2.20i$ for $\kappa=0.1$ 
[panel (b)] and $\Lambda= 1.4090$, $\lambda_3=-1.2632+2.25i$ for $\kappa=0.08$ [panel (c)]. When $\Lambda>|{\rm Re}(\lambda_3)|$, the residual $f(\Delta x)$ only has one root $\Delta x = L_x/2$, indicated by the vertical dashed line in panels (a) and (b), while for $\Lambda<|{\rm Re}(\lambda_3)|$, in panel (c), there are many equilibria. In these examples, we assumed that $g_1=-g_2$ and $\varphi=0$. The vertical axis is symmetric about zero and logarithmic, with a cutoff at $\pm10^{-3}$ [panel (a)] and $\pm10^{-21}$ [panels (b), (c)]}
    \label{fig:f_of_x}
\end{figure}

\begin{figure}
    \centering
    \includegraphics[width=0.5\linewidth]{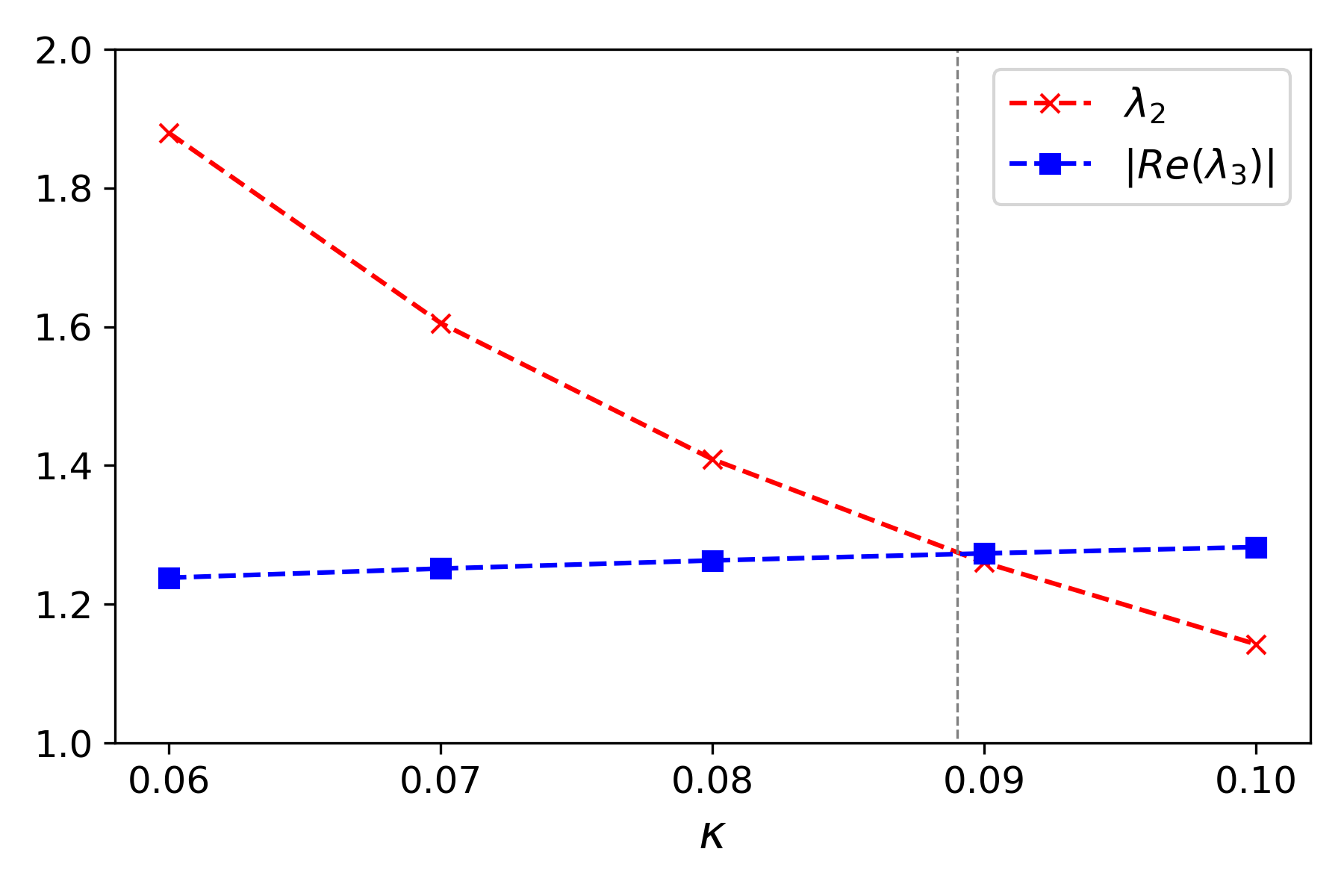}
    \caption{Leading spatial eigenvalue $\lambda_2$ (red crosses interpolated by red dashed line) and $|{\rm Re}(\lambda_3)|$ (blue squares interpolated by blue dashed line) as a function of $\kappa$, revealing a cross-over point at \textcolor{black}{$\kappa_c'\approx 0.089$}, denoted by the vertical gray dashed line. \textcolor{black}{The spatial eigenvalues shown here were computed with $c$ measured for a single pulse in $L_x=40$.}}
    \label{fig:lamb23_vs_kappa}
\end{figure}

\textcolor{black}{For a better understanding of the possible equilibrium values of $\Delta x$ when $\Lambda=\Lambda_r\in \mathbb{R}_+$, we subtract Eq.~(\ref{eq:red_a}) from Eq.~(\ref{eq:red_b}), with $c_1=c_2$, assuming $x_2> x_1$ without loss of generality, to obtain
\begin{equation}
    \frac{{\rm d}\Delta x}{{\rm d}t} = g_1 \textcolor{black}{\cos\varphi} \left[e^{-\Lambda_r (L_x-\Delta x) } - e^{-\Lambda_r \Delta x}\right] + g_2 \left[ \cos(\lambda_{3i}\Delta x + \varphi) e^{\lambda_{3r} \Delta x} - \cos(\lambda_{3i}(L_x-\Delta x) + \varphi)e^{\lambda_{3r} (L_x-\Delta x)} \right] \equiv f(\Delta x).\label{eq:dDelta_dt}
\end{equation}
Figure~\ref{fig:f_of_x} shows the function $f(\Delta x)$ for three cases in a domain of size $L_x=80$ (obtained using $g_1=-g_2$, $\varphi=0$, as in Fig.~\ref{fig:ode_dns_repulsion}). At $\kappa=0.95$ [panel (a)] and $\kappa=0.1$ [panel (b)], the function $f(\Delta x)$ only has one root, $\Delta x = L_x/2=40$. The significantly smaller values of $f(\Delta x)$ at $\kappa=0.1$ compared to $\kappa=0.95$ indicate a drastic slowing-down of the repulsion. In contrast, at $\kappa=0.08$ [panel (c)], there are oscillations leading to a large number of roots, corresponding to bound state separations, half of which are stable equilibria and the other half unstable.  The critical difference between the three cases is the relative size of $\Lambda$ compared with $|\lambda_{3r}|$ that controls which tail is more extended. 
For $\Lambda< |\lambda_{3r}|$ (monotonic tail longer than oscillatory tail), one finds numerically that the only root of $f(\Delta x)$ systematically occurs at $\Delta=L_x/2$, independently of $L_x$, and corresponds to a stable equilibrium, in agreement with Fig.~\ref{fig:ode_dns_repulsion}(b). However, this does not remain true for $\Lambda\geq |\lambda_{3r}|$ (oscillatory tail longer than monotonic tail), which is the case for weakly supercritical $\kappa\gtrsim \kappa_c$ (see Fig.~\ref{fig:spaeigcomp}), and/or $|g_2|\gg |g_1|$, in which case there are systematically multiple equilibria of  Eq.~(\ref{eq:reduced_model}) corresponding to multiple possible bound state separations. We note for completeness that for $|g_2| \gg |g_1|$ (a case which we did not observe in DNS) the function $f(\Delta x)$ has several roots even when $\Lambda$ is smaller than $|\lambda_{3r}|$. These findings complement the DNS results in Fig.~\ref{fig:40nonconv}, where convergence could not be reached due to limited simulation time, suggesting that bound states at finite separation form at $\kappa\gtrsim \kappa_c$, where $\Lambda\gtrsim |\lambda_{3r}|$. While longer DNS are needed to confirm these predictions, the model, if accurate, indicates that there is a second cross-over value of $\kappa$ somewhat larger than $\kappa_c$, corresponding to $\Lambda=\lambda_2=|\lambda_{3r}|$. Figure~\ref{fig:lamb23_vs_kappa} shows that this cross-over value of \textcolor{black}{$\kappa_c'$} is approximately $0.089$. \textcolor{black}{Based on this cross-over value $\kappa_c'$ (Fig.~\ref{fig:lamb23_vs_kappa}) obtained from the reduced-order model, we speculate that the final state will be a bound state when $\kappa<\kappa_c'$ (e.g., the cases $\kappa=0.06$ and $0.08$ in Fig.~\ref{fig:40nonconv}), while an equispaced state will be reached when $\kappa>\kappa_c'$ (e.g., the case $\kappa=0.1$ in Fig.~\ref{fig:40nonconv}).} We note that the observed repulsion in DNS and in the ordinary differential equation model aligns with the findings in \cite{kawahara1988pulse} for the generalized Kuramoto-Sivashinsky equation, where an interaction of one oscillatory tail and one monotonic tail was also found to result in repulsion of localized pulses. }

\subsection{Collision for $\kappa<\kappa_c$}
\begin{figure}
    \centering
    \includegraphics[width=0.4\linewidth]{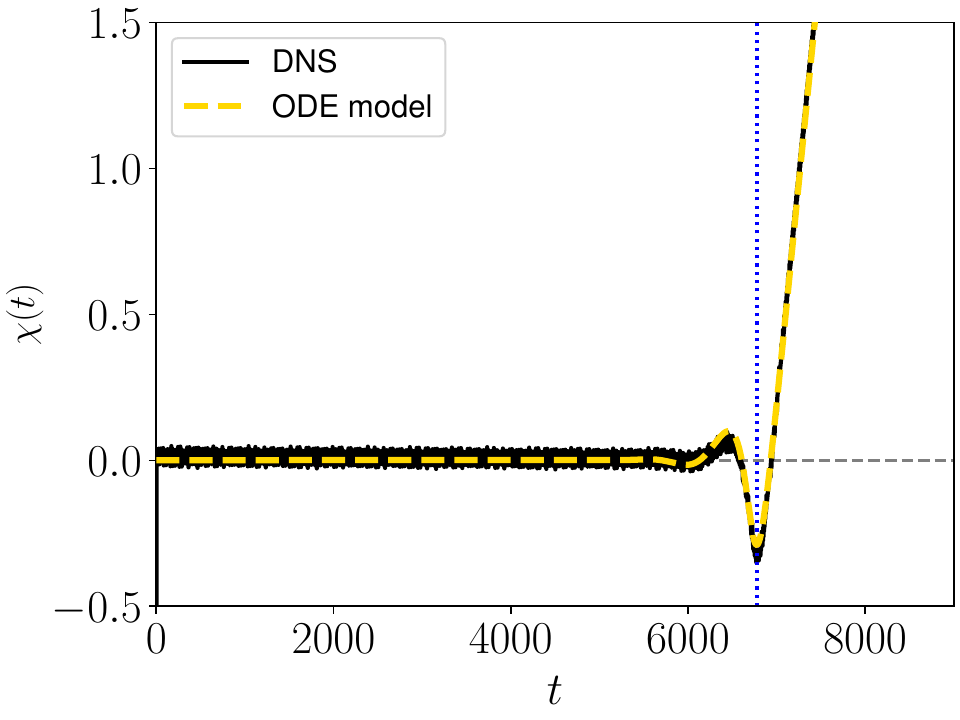}
    \caption{Compensated distance  $\chi(t)$ defined in Eq.~(\ref{eq:def_chi}) versus time from DNS (black line, noisy due to finite grid resolution) at $\kappa=0.03$ in a domain of size $L_x=80$ (Fig.~\ref{fig:collisions2}), and the reduced model (yellow dashed line) with $g_1=-1.62145867$, $g_2= 0.18617327$, $\varphi=1.85636433$, estimated by a fitting procedure with spatial eigenvalues $\Lambda= 1.5884+ 2.2528i$ and $\lambda_3=-1.5884+ 2.2528i$. The drift velocities are $c_1=-0.14165$ and $c_2=-0.14445$. At early times $\chi(t)\approx 0$ since interactions are weak. For reference, $\chi=0$ is indicated by the horizontal gray dashed line at all times. At $t\approx 6000$, deviations from $\chi =0$ indicate the onset of significant interactions of alternating sign between the two localized structures. The vertical blue dotted line indicates the time when the new three-pulse bound state is formed in the collision, after which the relative distance remains constant, leading to a linear increase in $\chi(t)$, cf.~Eq.~(\ref{eq:def_chi}). 
    }
    \label{fig:chi_collision}
\end{figure}

Next, we compare the reduced model with the DNS results for the chasing, inelastic collision between a single pulse and a two-pulse bound state, shown in Fig.~\ref{fig:collisions2}(a) at $\kappa=0.03$ (where $\Lambda=\lambda_1$) in a domain of size $L_x=80$, leading to the formation of a bound state of three pulses. We again initialize the reduced model with the same initial positions of the proximate maxima as in the DNS. To compare DNS and reduced model results quantitatively, we consider the deviation of the relative distance between proximate maxima from free propagation, namely 
\begin{equation}
\chi(t) \equiv \Delta x(t) - (c_2-c_1)t - \Delta x(t=0), \label{eq:def_chi}
\end{equation}
where we recall that $\Delta x(t)\equiv |x_2-x_1|$. When the two structures are located far apart from one another $\chi=0$, and nonzero values of $\chi$ are caused by interactions. Figure~\ref{fig:chi_collision} shows a comparison between $\chi(t)$ from DNS (black line, noisy due to finite grid resolution) and the reduced model (yellow dashed line). At early times $\chi(t)\approx 0$ since interactions are exponentially weak, leading to effectively free propagation. For reference $\chi=0$ is indicated by the horizontal gray dashed line at all times. At $t\approx 6000$, deviations from $\chi =0$ indicate the onset of significant interactions of alternating sign between the two localized structures: first, a small deviation towards negative $\chi$ can be seen, signifying attraction, followed by a larger-amplitude positive deviation in $\chi$, indicating repulsion, which is in turn followed by a larger-amplitude, negative deviation in $\chi$, implying renewed attraction. The evolution shown in Fig.~\ref{fig:chi_collision} closely resembles the results reported for collisions of localized structures in the Swift-Hohenberg equation \cite{raja2023collisions}. The vertical blue dotted line in Fig.~\ref{fig:chi_collision} indicates the time when the new three-pulse bound state is formed in the collision, after which the relative distance is constant, leading to a linear increase in $\chi(t)$, cf. Eq.~(\ref{eq:def_chi}).  Since $\chi$ is negative at that time, the collision is effectively attractive in the language proposed in \cite{raja2023collisions}: the formation of the new three-pulse bound state occurs more quickly due to the alternating attractive/repulsive interactions, increasing in strength as the two structures approach each other, than it would by free propagation. Finally, we note that the reduced model can be simplified in this case: the effect of the finite domain size is negligible, since at all times the proximate maxima are much closer to each other than $L_x/2$, and therefore the terms involving $L_x-\Delta x$ in Eq.~(\ref{eq:reduced_model}), which represent the interaction with periodic copies, are exponentially suppressed, a conclusion we have verified explicitly (not shown).

The successful quantitative description of the evolution of localized solutions of the system of partial differential equations in Eq.~(\ref{eq:perturbation_equations}), including repulsion and inelastic collisions, by the reduced system of ordinary differential equations in Eq.~(\ref{eq:reduced_model}) is indicative of the fact that the interactions between the localized states are to a large extent particlelike, with a nontrivial interaction potential determined by the linearized problem in Eq.~(\ref{eq:spatial_eigenvalue}). 

\section{Discussion and Conclusion}
\label{sec:conclusion}

In this work, we have investigated the dynamics of traveling spatially localized convective structures \textcolor{black}{(`traveling convectons')} in an inclined porous layer heated from below. The drift of the convectons, observed here for the first time in a porous medium, is a result of the asymmetry in the top and bottom temperature boundary conditions. For a fixed inclination angle and a fixed Rayleigh number, we conducted extensive DNS \textcolor{black}{(over 300 individual simulations)} for different values of the symmetry-breaking parameter $\kappa$, related to the Biot number, and different horizontal domain sizes $L_x$. In DNS with fixed domain size and varying $\kappa$, stable, traveling $n$-pulse states are typically observed in the weak symmetry-breaking regime ($\kappa \lesssim 0.1$). However, in sufficiently small domains $n$-pulse \textcolor{black}{traveling convectons} become unstable as $\kappa$ increases towards $\kappa=1$ (strong symmetry breaking). For traveling states consisting of more than one pulse, spatially localized structures can either form a bound state at a small separation or be evenly spread out across the finite computational domain. The drift velocity $c$ of traveling convectons is sensitive to $\kappa$ (as well as the domain size), exhibiting nontrivial, generically nonlinear relationships $c(\kappa)$. In small domains, the drift velocity $c$ was found to be positive (indicating an upslope travel direction) and monotonically increasing with $\kappa$, while in larger domains $c$ is sign-indefinite for localized structures consisting of a few pulses only. DNS with fixed $\kappa$ and increasing domain size revealed a strong impact of the domain size, which can be attributed to interactions between image structures in adjacent periodic domains. With strong symmetry breaking, multi-pulse localized structures can either decay directly to the conduction state or transition into traveling states with fewer pulses (partial annihilation).

To investigate the long-range interactions between localized structures mediated by their small-amplitude tails, we studied the spatial eigenvalues responsible for the tail profiles. We showed that the spatial eigenvalues which we computed numerically successfully predict the spatial growth rate and wave number of the leading and trailing tails of traveling localized structures in DNS, with accuracy that increases with increasing size of the computational domain. Moreover, the dominant spatial eigenvalue associated with the upslope tail (with a negative real part), which may be leading or trailing depending on the sign of $c$, remains complex for all parameter values, indicating that the upslope tail is always oscillatory, at least in the parameter regime analyzed here. The dominant spatial eigenvalue associated with the downslope tail (with a positive real part), which can likewise be leading or trailing, shows a competition between a pair of complex eigenvalues and a real eigenvalue, resulting in a transition from an oscillatory tail at small $\kappa$ to a monotonic tail at large $\kappa$. The spatial eigenvalues predict a critical threshold value $\kappa_c \approx 0.059$ for this transition in broad agreement with DNS observations. 

The discovery of this transition in the tail profiles sheds light on the interactions between localized structures observed in our DNS. We investigated the long-time dynamics of the interaction between localized pulses by measuring their spatial separation as a function of time. The results suggest that at $\kappa \leq 0.04$, which is in the parameter regime where tails are purely oscillatory, separation converges to a value much less than the largest possible separation in the domain, indicating the existence of bound states in a domain of size $L_x = 40$. This result aligns with the fact that interaction between oscillatory tails of localized structures is usually associated with the formation of bound states. We emphasize that the bound states are by no means unique: we expect a multiplicity of such states differing only in the interpulse separation. This separation is in turn determined by the locking of adjacent pulse tails. However, these states are not observed here and may be unstable.

On the other hand, for $\kappa > \kappa_c$ our DNS results for a traveling two-pulse structure confirm that for $\kappa > 0.2$ adjacent pulses repel one  
another, resulting ultimately in the largest possible separation in the domain ($\Delta x = 20$). The repulsive interaction resulting from the change in the tail profile for $\kappa>\kappa_c$ is not fully investigated numerically in this work. This is because, on the one hand, the interaction decreases rapidly with the separation between the two structures, and on the other hand, the drift velocity also decreases significantly in the weakly broken symmetry regime (i.e. $\kappa \lesssim 0.1$). As a result, very long simulation times are required to verify whether the final traveling localized structures are evolving towards a bound state or an equispaced state when the domain is large. So far, we have yet to observe definitive traveling equispaced states (or traveling bound states) in simulations with $0.04 <\kappa \leq 0.2$ in a $L_x = 40$ domain. For the same reason, in a $L_x = 80$ domain, we are able to verify that pulses become strictly equispaced only for $\kappa \geq 0.95$.
Complementing these DNS results obtained at the expense of a significant amount of computation time, we adapted a reduced-order model from \cite{raja2023collisions} and showed that it also accurately reproduces the interactions between localized structures in the present model, including pulse repulsion (for $\kappa\gg \kappa_c$) and the inelastic collision of bound states traveling at different velocities (for $\kappa<\kappa_c$). The reduced complexity of this model has provided significant insight into the dynamics, suggesting the existence of a second transition value of $\kappa$, with \textcolor{black}{$\kappa=\kappa_c'>\kappa_c$}. More precisely, the reduced model suggests that in the presence of one oscillatory tail and one monotonic tail and \textcolor{black}{$\kappa>\kappa_c'$}, the resulting repulsion of identical pulses always leads to an equispaced configuration at late times, while in the weakly supercritical regime \textcolor{black}{$\kappa_c<\kappa<\kappa_c'$} the reduced model indicates that bound states with a smaller separation than in the equispaced configuration may continue to form, although longer DNS are needed to test this prediction. Nevertheless, the good agreement between the reduced model and the solutions of the full fluid equations indicates that the interactions of localized convective structures are to a great extent particle-like.

A key aspect of this work was to gain insight into the effects of breaking a reflection symmetry, here the symmetry ${\cal R}$, and into the ensuing dynamics, but in other systems rotation \cite{ecke1992hopf} or throughflow \cite{buchel2000influence} play a similar role. In systems with periodic boundary conditions a spatially periodic state with a nonzero wave number generically begins to drift when the symmetry is broken, either externally as here or through spontaneous symmetry breaking, i.e., at a parity-breaking bifurcation. However, large domains can support multiple traveling states with distinct wave numbers and velocities, and one expects to find transitions between them as parameters are varied, cf.~\cite{knobloch1983}. These transitions may be hysteretic or continuous, but in each case occur via a branch of two-frequency states connecting the primary states \cite{Knobloch_1994}. These statements apply equally to spatially localized states. We conjecture that the Newton's cradle solutions (Figs.~\ref{fig:3p004} and \ref{fig:5p004}) are examples of stable two-frequency states of this type, connecting branches of single-frequency traveling states at either end of their existence interval. The fact that the two- and four-pulse traveling states are simultaneously stable provides evidence for a hysteretic transition between them. Evidently, numerical continuation techniques would be helpful to resolve some of these conjectures. We mention that disconnected asymmetric localized traveling states may be present even in systems with reflection symmetry, and accessed - as here - via a finite amplitude perturbation only~\cite{lo2017localized}.

It is evident that many aspects of the present problem remain to be explored. For weakly broken midplane reflection symmetry, the drift speed of localized structures can in many cases be predicted using asymptotic techniques, see e.g. \cite{raja2023collisions}, and adapting these techniques to the present system, where the symmetry-breaking parameter $\kappa$ only appears in the boundary conditions, will be an important next step. Furthermore, while the present work largely relies on DNS, numerical continuation and a stability analysis of spatially localized structures in this system would provide crucial additional insights. This would in particular help clarify to what extent this system follows typical bifurcation structures of spatially localized structures such as homoclinic snaking found in the Swift-Hohenberg equation \cite{PhysRevE.73.056211,knobloch2015spatial} and how this is impacted by the broken reflection symmetry ${\cal R}$. Finally, this study focused exclusively on the case $Ra=100$, $\phi=35^\circ$, and other choices of parameters, as well as a three-dimensional geometry, remain of interest. \textcolor{black}{The current 2D results can be viewed as describing transverse rolls where the roll axes are orthogonal to the shear flow arising from buoyancy. It is well known, however, that in 3D certain parameter regimes favor longitudinal rolls \citep{caltagirone1985solutions,wen2018inclined}, where the roll axes are aligned with the shear flow direction. As a result, it is reasonable to expect much richer flow physics will appear in 3D, much as occurs in 3D inclined Rayleigh-B\'enard convection \citep{reetz2020invariant1,reetz2020invariant2}.}
%reetz2020invariant_part1,reetz2020invariant_part2}.}

\begin{acknowledgments}
This work was supported by the National Science Foundation (Grants DMS-2009563, DMS-2308337 and OCE-2023541) and by the Deutsche Forschungsgemeinschaft (DFG Projektnummer 522026592). The computational resources for this project were provided by the NSF ACCESS program (project number: PHY230056), allowing us to utilize the Advanced Research Computing at the Johns Hopkins (ARCH) core facility (rockfish.jhu.edu), which is supported by the National Science Foundation under Grant No. OAC 1920103, and the Purdue Anvil CPU cluster \cite{song2022anvil}. C.L. acknowledges the support from UConn Research Excellence Program and UConn Quantum Innovation Seed Grants. 
\end{acknowledgments}

% \appendix

% % The \nocite command causes all entries in a bibliography to be printed out
% % whether or not they are actually referenced in the text. This is appropriate
% % for the sample file to show the different styles of references, but authors
% % most likely will not want to use it.
% \nocite{*}

\bibliography{apssamp}% Produces the bibliography via BibTeX.

\end{document}